\documentclass[preprint,aps]{revtex4}
\usepackage{graphicx,slashed,color}
\begin{document}
%\preprint{}
\draft
\title{Stueckelberg Mechanism and Chiral Lagrangian for $Z'$ Boson}

\author{Ying Zhang$^1$, Shun-Zhi Wang$^2$, Qing Wang$^3$}

\address{
    School of Science, Xi'an Jiaotong University, Xi'an, 710049, P.R.China$^1$\footnote{hepzhy@mail.xjtu.edu.cn}\\
    Department of Physics, Tsinghua University, Beijing, 100084, P.R.China$^1$~$^2$\footnote{wsz04@mails.tsinghua.edu.cn}
    $^3$\footnote{wangq@mail.tsinghua.edu.cn}\\
    Center for High Energy Physics, Tsinghua University, Beijing 100084, P.R.China$^3$}

\date{Mar 8, 2008}

\begin{abstract}
Traditional Stueckelberg Mechanism is shown equivalent to set up a
gauged $U(1)$ chiral Lagrangian and fix special gauge. With this
mechanism, the original electro-weak chiral Lagrangian is inlarged
by including an extra $U(1)$ symmetry to represent physics for $Z'$
boson. We build up complete list of electro-weak chiral Lagrangian
up to order of $p^4$ including $Z'$ and higgs bosons.  The most
general mixing among neutral gauge bosons is diagonalized completely
and the connections among these operators to triple, quartic
couplings involving $Z'$ boson and to that in traditional
electro-weak chiral Lagrangian are made.

\bigskip
PACS number(s): 11.10.Ef; 11.10.Lm; 12.15.-y; 12.15.Mm; 12.39.Fe;
12.60.Cn
\end{abstract}
\maketitle

%%%%%%%%%%%%%%%%%%%%%%%%%%%%%%%%%%%%%%%%%%%%%%
\section{Introduction}

Seventy years before, C.G. Stueckelberg \cite{Origin} introduced a
scalar into massive abelian vector theory without violation of gauge
symmetry and renormalizability. Since then, people used to apply
this mechanism to describe massive photon.  Beyond that, many other
applications also emerged, such as those to
SM\cite{KorsNathPLB}\cite{Z'LHC}, MSSM\cite{KorsNathinJHEP2005},
string\cite{Ramond} and extra dimension\cite{WrapedExtraDim}, etc.
The latest review was given by Ref.\cite{StueckRuegg}. This
mechanism in literature was seen as a scheme to replace Higgs
mechanism for broken $U(1)$ gauge theory \cite{FirstReplaceHiggs} in
the sense that it does not need Higgs particle. Among various
applications, we are interested in this work in investigating
physics of $Z'$ boson. On the one hand, as a heavy undiscovered new
vector particle in the minimal extension of SM, $Z'$ will probably
be the particle easest to test in future collider experiments and
 plays important role in various new physics models, such as low
energy models induced from GUT and
 SUSY \cite{DittmarPLB2004}\cite{Leike1999}\cite{HewettRizzoZprime}\cite{FranziniPRD1987},
left-right symmetric models\cite{DittmarPLB2004}, little Higgs
models\cite{LittleHiggsZprimePLB2001} and extra dimension
models\cite{AntoniadisKK}\cite{GizzoKK}, etc; on the other hand,
Stueckelberg mechanism provides us a special method to introduce
abelian massive vector into theory gauge invariantly. With this
mechanism, we can simply add $Z'$ boson to SM and discuss
corresponding physics \cite{KorsNathPLB}. However, the traditional
Stueckelberg mechanism only deals with lowest dimension term related
to vector boson mass and leads typical mixing term between scalar
particle and gauge boson, which  does not include those more complex
high dimension operators. As a consequence, this approach lost
generality in the sense that operator involving $Z'$ boson through
Stueckelberg mechanism is that with lowest dimension which
 represents a special kind of $Z'$ interaction. Though this operator
 plays the most important role in low energy region, it is not general
enough when we approach to TeV energy region where effects of high
dimension operators will emerge. These high dimensional operators,
most of them are non-renormalizable, are effective description of
underlying new physics dynamics. Adding in these non-renormalizable
high dimension operators into theory is a necessary step when we
want to go beyond SM to investigate new physics model independently.
This  requirement leads to the generalization of the traditional
Stueckelberg mechanism by including high dimension operators into
theory so that general $Z'$ interactions may be covered as much  as
possible. With non-renormalizable operators included in,
renormalizability of original theory is lost and is replaced by a
generalized version of renormalization for effective field
theory\cite{Weinberg}.

 There are
two ways to systematically describe general effective interactions
among particles in SM: namely, linear and nonlinear realizations of
SM symmetry $SU(2)_L\otimes U(1)_Y\rightarrow U(1)_{em}$. Within
linear realization, we just add in high dimension operators into SM
\cite{Buchmuller86,hagiwara96,Gonzalez-Garcia99,ARZT95}. While in
nonlinear realization, we start from electroweak chiral Lagrangian
(EWCL)\cite{longhitano80,Appelquist80,Appelquist93} which is the
most general description for SM fields except Higgs. This EWCL was
generalized to extended electroweak chiral Lagrangian (EEWCL) by
adding in original EWCL a singlet Higgs field\cite{WangLiMing} to
keep unitarity of the theory \cite{unitarity}. Though mathematical
equivalence between two descriptions was shown in \cite{WangLiMing},
linear realization is suitable for discussion of light Higgs, while
nonlinear realization can be applied to investigate either light or
heavy Higgs. Due to this generality for EEWCL, we use nonlinear
realization in this paper. In fact, we will show that Stueckelberg
mechanism is  equivalent to chiral Lagrangian for
 $U(1)$ gauge field plus special choice of gauge fixing term.
This equivalence enable us to further understand the
non-renormalizability for  Stueckelberg mechanism when we try to
generalize it to non-abelian gauge field system and base our whole
discussion on the nonlinear realization of SM symmetry. With the
equivalence of Stueckelberg mechanism and $U(1)$ chiral Lagrangian,
the generalization of traditional Stueckelberg mechanism become
obvious:  we just extend EEWCL with an extra $U(1)$ gauge symmetry
and write down all possible high dimension interaction terms. To
make particle content in our discussion close to low energy particle
spectrum already discovered in experiment, except Higgs and $Z'$
bosons, we do not involve any other new undiscovered particles in
our theory. Higgs particle in this work only plays a passive role
and we mainly focus our attention on $Z'$ interactions.

This paper is organized as follows. Sec.II is the proof for the
equivalence of traditional Stueckelberg mechanism and $U(1)$ chiral
Lagrangian and discussion of its nonabelian generalization. In
Sec.III, we generalize original EEWCL to include $Z'$ boson and
write down the bosonic part of Lagrangian up to order of $p^4$. From
this Lagrangian, we obtain the most general mixing for neutral gauge
bosons. Then we completely diagonalize and discuss the mixing. In
Sec.IV, We build up the connections of these operators to triple ,
quartic couplings involving $Z'$ boson and traditional electro-weak
chiral Lagrangian. The summary is given in Sec.IV.

%%%%%%%%%%%%%%%%%%%%%%%%%%%%%%%%%%%%%%%%%%%%%%%
\section{Equivalence between Stueckelberg Mechanism and chiral Lagrangian}

 Now, let us review Stueckelberg mechanism.
The most simple Stueckelberg Lagrangian for massive vector $A_\mu$
can be written as
\begin{eqnarray}
\mathcal{L}_{Stueck}&=&-\frac{1}{4}F_{\mu\nu}F^{\mu\nu}+\frac{m^2}{2}(A_\mu-\frac{1}{m}\partial_\mu\sigma)^2\;,
\label{LagrangianStueck}
\end{eqnarray}
with obvious mass term ${m^2}A_\mu^2/2$. Under $U(1)$ gauge
transformation $A_\mu\rightarrow A_\mu+\partial_\mu\epsilon$,
$\sigma\rightarrow\sigma+m\epsilon$, the Lagrangian is invariant.
Adding a gauge fixing term
\begin{eqnarray}
\mathcal{L}_{GF}&=&-\frac{1}{2\xi}(\partial_\mu A^\mu+\xi
m\sigma)^2\label{gfix}
\end{eqnarray}
into the Lagrangian, the total Lagrangian is the sum of Stueckelberg
Lagrangian $\mathcal{L}_{Stueck}$ and gauge fixing term
$\mathcal{L}_{GF}$
\begin{eqnarray}
\mathcal{L}_{total}&=& -\frac{1}{4g^2}F_{\mu\nu}F^{\mu\nu}
    +\frac{m^2}{2}A_\mu A^\mu
    -\frac{1}{2\xi}(\partial_\mu A^\mu)^2
    +\frac{1}{2}(\partial_\mu\sigma)^2
    -\frac{\xi}{2} m^2\sigma^2.
\end{eqnarray}
Mixing term $\sigma\partial_{\mu}A^\mu$ appeared in
$\mathcal{L}_{GF}$ cancels the same term in $\mathcal{L}_{Stueck}$.
This leads to the decoupling of auxiliary scalar $\sigma$ and vector
field $A_\mu$. The unphysical $\sigma$ is given a mass proportional
to random parameter $\sqrt{\xi}$, which means $\sigma$ is unphysical
field and have no any influence on vector field $A_\mu$. So
traditional Stueckelberg mechanism include two parts. One is
extension of standard mass term of $U(1)$ gauge boson through term
mixing with differential of scalar field. This part, we will show,
is equivalent to gauged $U(1)$ chiral Lagrangian. The other is
choice of special gauge fixing term to cancel mixing between scalar
and gauge boson.

Now we prove the assertion that the first part of traditional
Stueckelberg mechanism is equivalent to gauged $U(1)$ chiral
Lagrangian. We change $\sigma$ field  by introducing an unitary
phase angle field $U$ as
\begin{eqnarray}
U(x)\equiv e^{i\frac{\sigma(x)}{m}}\;.
\end{eqnarray}
Under $U(1)$ gauge transformation, it transforms as $U\rightarrow
e^{i\epsilon}U$. We can construct covariant derivative for $U$ as
\begin{eqnarray}
D_{\mu}U(x)\equiv[\partial_\mu-iA_\mu(x)]U(x)=iU(x)\bigg[\frac{1}{m}[\partial_\mu\sigma(x)]-A_\mu(x)\bigg]\;.
\end{eqnarray}
With this covariant derivative, we can rewrite
(\ref{LagrangianStueck}) in terms of $U$ field as
\begin{eqnarray}
\mathcal{L}_{Stueck}&=&-\frac{1}{4g^2}F_{\mu\nu}F^{\mu\nu}+\frac{m^2}{2}(D^{\mu}U)^\dag(D_{\mu}U)\;,
\label{LagrangianStueck1}
\end{eqnarray}
which is standard lowest $p^2$ order chiral Lagrangian (gauged
nonlinear $\sigma$ model) for $U(1)$ gauge field as long as we
identify $m$ with goldstone decay constant $f$. Here $\sigma$ plays
the role of goldstone boson which, in terms of Higgs mechanism, will
be eaten out by gauge field $A_\mu$ to become its longitudinal part
after symmetry breaking. Broken $U(1)$ symmetry is explicitly seen
through unitary gauge $U=1$ ( or taking vacuum).

In terms of our $U$ field representation, gauge fixing term
(\ref{gfix}) can be written as
\begin{eqnarray}
\mathcal{L}_{GF}&=&-\frac{1}{2\xi}(\partial_\mu A^\mu-i\xi m^2\ln
U)^2\;,\label{gfix1}
\end{eqnarray}
which can cancel the mixing term between $A_\mu$ and $\sigma$ and
make $\sigma$ becoming free field.

Above equivalence between Stueckelberg mechanism and gauged $U(1)$
chiral Lagrangian can be seen as an alternative statement for the
distinction of the Stueckelberg  and the Higgs mechanisms for which
conventional understanding relies on the existance of a Higgs
particle \cite{KorsNathinJHEP2005}. Now chiral Lagrangian is a
formalism constructed by gauge field and corresponding goldstone
boson, it does not need Higgs field and therefore in this sense is
the same as Stueckelberg mechanism. In fact, this equivalence was
pointed out in an alternative way in Ref.\cite{equivalence}. With
this equivalence, applications of Stueckelberg mechanism can be
realized in terms of standard formalism of chiral Lagrangian. One
possible application is to consider effects from high dimension
operators which as mentioned in last section may reflect more
complex and general interactions among $Z'$ boson and SM particles.
This will be discussed in next section. Another direction of
application is to generalize $U(1)$ to nonabelian gauge symmetry. In
the following part of this section, we take a simplest nonabelian
generalization by considering following symmetry breaking
realization $SU(2)_L\otimes SU(2)_R\rightarrow SU(2)_D$ with
$2\times 2$ unitary matrix field $\tilde{U}$ defined as
\begin{eqnarray}
\tilde{U}(x)\equiv
e^{\frac{i}{m}\tilde{\sigma}_i(x)\tau_i}=m[\sqrt{1-\frac{\Sigma^2(x)}{m^2}}+i\frac{\Sigma_i(x)\tau_i}{m}]
\hspace{2cm}\Sigma_i(x)\equiv
\frac{m\tilde{\sigma}_i(x)}{\sqrt{\tilde{\sigma}^2(x)}}\sin\frac{\sqrt{\tilde{\sigma}^2(x)}}{m}\;,\label{tildeUdef}
\end{eqnarray}
where $\tau_i,~i=1,2,3$ are Pauli matrices,  and $\tilde{\sigma}_i$
are three goldstone bosons generated from $SU(2)_L\otimes
SU(2)_R\rightarrow SU(2)_D$ through Goldstone theorem. The
$SU(2)_L\otimes SU(2)_R$ gauge transformation is
\begin{eqnarray}
\tilde{U}(x)\rightarrow
\tilde{V}_R(x)\tilde{U}(x)\tilde{V}_L^\dag(x)\label{SU2}
\end{eqnarray}
in which $\tilde{V}_R$ and $\tilde{V}_L$ are $SU(2)_R$ and $SU(2)_L$
group elements respectively.

Note that if we return back from (\ref{SU2}) to our original abelian
situation, $U$ field will transform as
\begin{eqnarray}
U(x)\rightarrow V_R(x)U(x)V_L^\dag(x)=
V_R(x)V_L^\dag(x)U(x)\;,\label{U1}
\end{eqnarray}
where $V_L=e^{i\epsilon_L}$ and $V_R=e^{i\epsilon_R}$ is $U(1)_L$
and $U(1)_R$ group element, respectively. Consider $U(1)_L\otimes
U(1)_R=U(1)_D\otimes U(1)$, with $V_D=e^{i(\epsilon_R+\epsilon_L)}$
and $V_A=e^{i(\epsilon_R-\epsilon_L)}=V_RV_L^\dag$ being
corresponding $U(1)_D$ and $U(1)$ group elements respectively. From
(\ref{U1}), it is easy to see that $U$ field is invariant under
$U(1)_D$ transformation and therefore $U(1)_D$ is a trivial symmetry
for Lagrangian (\ref{LagrangianStueck1}). With this trivial $U(1)_D$
symmetry included in (\ref{LagrangianStueck1}), the symmetry
realization pattern for original Stueckelberg Lagrangian become
$U(1)_L\otimes U_R(1)\rightarrow U(1)_D$. With this form of ablelian
symmetry realization for  original Stueckelberg Lagrangian, our
nonabelian generalization of $SU(2)_L\otimes SU(2)_R\rightarrow
SU(2)_D$ become obvious. The only difference from abelian case is
that the left unbroken symmetry $SU(2)_D$ is not a trivial symmetry
in the sense that $\tilde{U}$ is not invariant under its
transformations.

Now we write down the Stueckelberg Lagrangian for $SU(2)_L\otimes
SU(2)_R\rightarrow SU(2)_D$,
\begin{eqnarray}
\mathcal{L}_{Stueck-SU(2)}&=&-\frac{1}{4g^2_L}F_{L,i}^{\mu\nu}F_{L,i}^{\mu\nu}-\frac{1}{4g^2_R}F_{R,i}^{\mu\nu}F_{R,i}^{\mu\nu}+\frac{m^2}{4}\mathrm{tr}[(D^{\mu}\tilde{U})^\dag
(D_{\mu}\tilde{U})]\;, \label{LagrangianStueck2}
\end{eqnarray}
with
\begin{eqnarray}
&&\hspace{-0.5cm}D^\mu\tilde{U}\equiv\partial^\mu\tilde{U}-i\frac{\tau_i}{2}(\tilde{V}_i^\mu+\tilde{A}_i^\mu)\tilde{U}
+i\tilde{U}\frac{\tau_i}{2}(\tilde{V}_i^\mu-\tilde{A}_i^\mu)\label{coDerDef}\\
&&\hspace{-0.5cm}F_{^R_L,i}^{\mu\nu}\frac{\tau_i}{2}=\partial^\mu(\tilde{V}^\nu\pm\tilde{A}^\nu)-\partial^\nu(\tilde{V}^\nu\pm\tilde{A}_i^\mu)
-i[\tilde{V}^\mu\pm\tilde{A}^\mu,\tilde{V}^\nu\pm\tilde{A}^\nu]\hspace{1cm}\tilde{V}^\mu\equiv\tilde{V}^\mu_i\frac{\tau_i}{2}
\hspace{1cm}\tilde{A}^\mu\equiv\tilde{A}^\mu_i\frac{\tau_i}{2}\;,\nonumber
\end{eqnarray}
where $\tilde{V}_i^\mu~~,i=1,2,3$ are $SU(2)_D$ gauge fields and
$\tilde{A}_i^\mu~~,i=1,2,3$ are $SU(2)_L\otimes SU(2)_R/SU(2)_D$
axial gauge fields. In unitary gauge, the third term of r.h.s. of
(\ref{LagrangianStueck2}) becomes mass term
$\frac{1}{2}m^2\tilde{A}^2$ of the axial gauge boson field
$\tilde{A}$. Due to unbroken symmetry $SU(2)_D$, corresponding gauge
fields $\tilde{V}_i^\mu~~,i=1,2,3$ remain massless.

In terms of fields $\Sigma_i$ which is already expressed as function
of $\tilde{\sigma}_i$ in (\ref{tildeUdef}), covariant derivative
(\ref{coDerDef}) now is
\begin{eqnarray}
D^\mu\tilde{U}=[-\frac{\Sigma_i}{m\sqrt{1-\frac{\Sigma^2}{m^2}}}+i\frac{\tau_i}{m}]\partial^\mu\Sigma_i+\frac{1}{m}
\tilde{A}_i^\mu[\Sigma_i-i\tau_i\sqrt{1-\frac{\Sigma^2}{m^2}}~]+
\frac{i}{m}\tilde{V}_i^\mu\Sigma_j\epsilon_{ijk}\tau_k\;.
\end{eqnarray}
With it, (\ref{LagrangianStueck2}) become
\begin{eqnarray}
\mathcal{L}_{Stueck-SU(2)}&=&-\frac{1}{4g^2_L}F_{L,i}^{\mu\nu}F_{L,i}^{\mu\nu}-\frac{1}{4g^2_R}F_{R,i}^{\mu\nu}F_{R,i}^{\mu\nu}
\label{LagrangianStueck3}\\
&&+\frac{1}{2}\left(-\frac{\Sigma_i}{\sqrt{1-\frac{\Sigma^2}{m^2}}}\partial^\mu\Sigma_i+\tilde{A}_i^\mu\Sigma_i\right)
\left(-\frac{\Sigma_{i'}}{\sqrt{1-\frac{\Sigma^2}{m^2}}}\partial_\mu\Sigma_{i'}+\tilde{A}_{i',\mu}\Sigma_{i'}\right)\nonumber\\
&&+\frac{1}{2}\left(\partial^\mu\Sigma_i-\tilde{A}_i^\mu\sqrt{1-\frac{\Sigma^2}{m^2}}+\tilde{V}_j^\mu\Sigma_k\epsilon_{ijk}\right)
\left(\partial_\mu\Sigma_i-\tilde{A}_{i,\mu}\sqrt{1-\frac{\Sigma^2}{m^2}}+\tilde{V}_{j'}^\mu\Sigma_{k'}\epsilon_{ij'k'}\right)\;.
\nonumber
\end{eqnarray}
We find that not only the terms linear in gauge fields
$\tilde{V}_i^\mu$ and $\tilde{A}_i^\mu$ mix with $\Sigma_j$ fields,
but the terms bilinear in gauge fields also  mix with $\Sigma_j$
fields which is the general feature for non-abelian gauged nonlinear
$\sigma$ model. This is not like the case of original abelian gauge
field, where terms bilinear in gauge fields do not mix with
$\Sigma_j$ fields. This feature makes it impossible to use gauge
fixing term to cancel mixing among gauge fields and goldstone
fields. Further nonabelian effects cause very complex dependence on
goldstone fields which make theory non-renormalizable. This example
explicitly shows why generalization of Stueckelberg mechanism to
non-abelian case can not cancel mixing among scalars and gauge
fields and then cause a coupled non-renormalizable theory.
%%%%%%%%%%%%%%%%%%%%%%%%%%%%%%%%%%%%%%%%%%%%%%%%%%%%%%%%%
\section{Generalized Stueckelberg Mechanism and EEWCL for $Z'$ boson}

As mentioned in Sec.I, nonlinear realized effective field theory
EEWCL is already worked out by one of us in Ref.\cite{WangLiMing}.
Although this EEWCL only involve boson fields in SM, it's enough for
our interests. In this section we are going to generalize it to
include in $Z'$ boson. The symmetry realization pattern is then
generalized from original $SU(2)_L\otimes U(1)_Y\rightarrow
U(1)_{em}$ to $SU(2)_L\otimes U(1)_Y\otimes U(1)_{Z'}\rightarrow
U(1)_{em}$. From equivalence between Stueckelberg mechanism and
chiral Lagrangian discussed in last section, to apply generalized
Stueckelberg mechanism to $Z'$ boson for EEWCL is equivalent to add
into EEWCL a phase degree of freedom representing goldstone boson
eaten out by $Z'$ and then gauging in $Z'$ gauge field. We insert
this goldstone boson degree of freedom by enlarging original two by
two unimodular matrix $U$ field with an extra $U(1)$ phase factor,
The new two by two field will be denoted by $\hat{U}$. The
difference between $U$ and $\hat{U}$ is that $U$ is unimodular which
satisfies constraint det$U=1$ while $\hat{U}$ does not. Relaxing
this unimodular constraint allows an extra $U(1)$ phase in $U$ field
which now is identified with mixture of goldstone bosons for $Z$ and
$Z'$. We define the covariant derivative as
\begin{eqnarray}
D_\mu\hat{U}=\partial_\mu\hat{U}+igW_\mu\hat{U}-i\hat{U}\frac{\tau_3}{2}g'B_\mu
-i\hat{U}(\tilde{g}'B_\mu+g^{\prime\prime}X_\mu)I.\label{DUdef}
\end{eqnarray}
where, $W_\mu\equiv\frac{\tau_i}{2}W^i_\mu$,  $B_{\mu\nu}$, $X_\mu$
are $SU(2)_L$, $U(1)_Y$ and $U(1)_{Z'}$ gauge fields respectively.
The reason to use $X$ instead of $Z'$ to label the $U(1)_{Z'}$ gauge
field is due to the fact that there exists mixing among neutral
gauge bosons. We denote $Z'$ as the $U(1)_{Z'}$ gauge field after
diagonaliztion. In (\ref{DUdef}), the new term beyond original
covariant derivative given in Ref.\cite{Appelquist93} is
proportional to the linear combination of gauge fields $B_\mu$ and
$X_\mu$ with different coefficients $\tilde{g}'$ and
$g^{\prime\prime}$. Different choice of these coefficients will
results in different $Z'$ interactions and typical $Z'$ dynamics
from non-traditional Stueckelberg mechanism usually take
$\tilde{g}'=0$. Later, we will discuss this issue in more detail.

 The full bosonic part Lagrangian up to order of
$p^4$ is
\begin{eqnarray}
\mathcal{L}_{Stueck-SU(2)_L\otimes U(1)_Y\otimes
U(1)_{Z'}\rightarrow
U(1)_{em}}=\mathcal{L}_0+\mathcal{L}_2+\mathcal{L}_4\;,
\end{eqnarray}
with $p^0$ and $p^2$ order Lagrangian $\mathcal{L}_0$ and
$\mathcal{L}_2$ being
\begin{eqnarray}
\mathcal{L}_0&=&-V(h)\;,\\
\mathcal{L}_2 &=&\frac{1}{2}(\partial_\mu h)^2
-\frac{1}{4}f^2\mathrm{tr}[\hat{V}_\mu\hat{V}^\mu]
+\frac{1}{4}\beta_1f^2\mathrm{tr}[T\hat{V}_\mu]\mathrm{tr}[T\hat{V}^\mu]
+\frac{1}{4}\beta_2f^2\mathrm{tr}[\hat{V}_\mu]\mathrm{tr}[T\hat{V}^\mu]\nonumber\\
&&+\frac{1}{4}\beta_3f^2\mathrm{tr}[\hat{V}_\mu]\mathrm{tr}[\hat{V}^\mu]
+\beta_4f(\partial^{\mu}h)\mathrm{tr}[\hat{V}_\mu]\;,\label{L2}
\end{eqnarray}
where $T\equiv\hat{U}\tau_3\hat{U}^\dag$ and $\hat{V}_\mu\equiv
(\hat{D}_\mu\hat{U})\hat{U}^\dag$. Here we treat higgs field $h$ as
$p^0$ order. All coefficients $f,\beta_1,\beta_2,\beta_3,\beta_4$
are functions of higgs field $h$. $p^4$ order Lagrangian
$\mathcal{L}_4$ can be decomposed into four parts
\begin{eqnarray}
\mathcal{L}_4&=&\mathcal{L}_K+\mathcal{L}_B+\mathcal{L}_H+\mathcal{L}_A
\end{eqnarray}
in which kinetic part $\mathcal{L}_K$ is
\begin{eqnarray}
\mathcal{L}_K&=&
    -\frac{1}{4}B_{\mu\nu}B^{\mu\nu}
    -\frac{1}{2}\mathrm{tr}[W_{\mu\nu}W^{\mu\nu}]
    -\frac{1}{4}X_{\mu\nu}X^{\mu\nu}\;.
\end{eqnarray}
Bosonic part without differential of higgs field $\mathcal{L}_B$ is
\begin{eqnarray}
\mathcal{L}_B
&=&\frac{1}{2}\alpha_1gg'B_{\mu\nu}\mathrm{tr}[TW^{\mu\nu}]
+\frac{i}{2}\alpha_2g'B_{\mu\nu}\mathrm{tr}[T[\hat{V}^\mu,\hat{V}^\nu]]
+i\alpha_3g\mathrm{tr}[W^{\mu\nu}[\hat{V}^\mu,\hat{V}^\nu]]\nonumber\\
&&+\alpha_4\mathrm{tr}[\hat{V}_\mu\hat{V}_\nu]\mathrm{tr}[\hat{V}^\mu\hat{V}^\nu]
+\alpha_5\mathrm{tr}[\hat{V}_\mu\hat{V}^\mu]\mathrm{tr}[\hat{V}^\nu\hat{V}_\nu]
+\alpha_6\mathrm{tr}[\hat{V}_\mu\hat{V}_\nu]\mathrm{tr}[T\hat{V}^\mu]\mathrm{tr}[T\hat{V}^\nu]\nonumber\\
&&+\alpha_7\mathrm{tr}[\hat{V}_\mu\hat{V}^\mu]\mathrm{tr}[T\hat{V}_\nu]\mathrm{tr}[T\hat{V}^\nu]
+\frac{1}{4}\alpha_8g^2\mathrm{tr}[TW_{\mu\nu}]\mathrm{tr}[TW^{\mu\nu}]
+\frac{i}{2}\alpha_9 g\mathrm{tr}[TW^{\mu\nu}]\mathrm{tr}[T[\hat{V}_\mu,\hat{V}_\nu]]\nonumber\\
&&+\frac{1}{2}\alpha_{10}\mathrm{tr}[T\hat{V}^\mu]\mathrm{tr}[T\hat{V}^\nu]\mathrm{tr}[T\hat{V}_\mu]\mathrm{tr}[T\hat{V}_\nu]
+\alpha_{11}g\epsilon^{\mu\nu\rho\lambda}\mathrm{tr}[T\hat{V}_\mu]\mathrm{tr}[\hat{V}_\nu
W_{\rho\lambda}]\nonumber\\
&&+\alpha_{12}g\mathrm{tr}[T\hat{V}^\mu]\mathrm{tr}[\hat{V}^\nu
W_{\mu\nu}]
+\alpha_{13}gg'\epsilon^{\mu\nu\rho\lambda}B_{\mu\nu}\mathrm{tr}[TW_{\rho\lambda}]
+\alpha_{14}g^2\epsilon^{\mu\nu\rho\lambda}\mathrm{tr}[TW_{\mu\nu}]\mathrm{tr}[TW_{\rho\lambda}]\nonumber\\
&&+\alpha_{15}\mathrm{tr}[\hat{V}_\mu]\mathrm{tr}[T\hat{V}^\mu]\mathrm{tr}[T\hat{V}_\nu]\mathrm{tr}[T\hat{V}^\nu]
+\alpha_{16}\mathrm{tr}[\hat{V}_\mu]\mathrm{tr}[T\hat{V}^\mu]\mathrm{tr}[\hat{V}_\nu\hat{V}^\nu]
+\alpha_{17}\mathrm{tr}[\hat{V}_\mu]\mathrm{tr}[T\hat{V}_\nu]\mathrm{tr}[\hat{V}^\mu\hat{V}^\nu]\nonumber\\
&&+\alpha_{18}\mathrm{tr}[\hat{V}_\mu]\mathrm{tr}[\hat{V}_\nu]\mathrm{tr}[T\hat{V}^\mu]\mathrm{tr}[T\hat{V}^\nu]
+\alpha_{19}\mathrm{tr}[\hat{V}_\mu]\mathrm{tr}[\hat{V}_\nu]\mathrm{tr}[\hat{V}^\mu\hat{V}^\nu]
+\alpha_{20}\mathrm{tr}[\hat{V}_\mu]\mathrm{tr}[\hat{V}^\mu]\mathrm{tr}[T\hat{V}_\nu]\mathrm{tr}[T\hat{V}^\nu]\nonumber\\
&&+\alpha_{21}\mathrm{tr}[\hat{V}_\mu]\mathrm{tr}[\hat{V}^\mu]\mathrm{tr}[\hat{V}_\nu\hat{V}^\nu]
+\alpha_{22}\mathrm{tr}[\hat{V}_\mu]\mathrm{tr}[\hat{V}^\mu]\mathrm{tr}[\hat{V}_\nu]\mathrm{tr}[T\hat{V}^\nu]
+\alpha_{23}\mathrm{tr}[\hat{V}_\mu]\mathrm{tr}[\hat{V}_\nu]\mathrm{tr}[\hat{V}^\mu]\mathrm{tr}[\hat{V}^\nu]\nonumber\\
&&+gg^{\prime\prime}\alpha_{24}X_{\mu\nu}\mathrm{tr}[TW^{\mu\nu}]
+g'g^{\prime\prime}\alpha_{25}B_{\mu\nu}X^{\mu\nu}
+\alpha_{26}\epsilon^{\mu\nu\rho\lambda}\mathrm{tr}[\hat{V}_\mu]\mathrm{tr}[T\hat{V}_\nu]\mathrm{tr}[T[\hat{V}_\rho,\hat{V}_\lambda]]\nonumber\\
&&+ig'\alpha_{27}\epsilon^{\mu\nu\rho\lambda}\mathrm{tr}[\hat{V}_\mu]\mathrm{tr}[T\hat{V}_\nu]B_{\rho\lambda}
+ig\alpha_{28}\epsilon^{\mu\nu\rho\lambda}\mathrm{tr}[\hat{V}_\mu]\mathrm{tr}[T\hat{V}_\nu]\mathrm{tr}[TW_{\rho\lambda}]\nonumber\\
&&+g\alpha_{29}\epsilon^{\mu\nu\rho\lambda}\mathrm{tr}[\hat{V}_\mu]\mathrm{tr}[\hat{V}_\nu
W_{\rho\lambda}]+ig^{\prime\prime}\alpha_{30}\epsilon^{\mu\nu\rho\lambda}X_{\mu\nu}\mathrm{tr}[T[\hat{V}_\rho,\hat{V}_\lambda]]
+ig^{\prime\prime}\alpha_{31}X_{\mu\nu}\mathrm{tr}[T[\hat{V}^\mu,\hat{V}^\nu]]\nonumber\\
&&+g^{\prime\prime}\alpha_{32}\epsilon^{\mu\nu\rho\lambda}\mathrm{tr}[\hat{V}_\mu]\mathrm{tr}[T\hat{V}_\nu]X_{\rho\lambda}
+\alpha_{33}\mathrm{tr}[\hat{V}_\mu]\mathrm{tr}[T\hat{V}_\nu]\mathrm{tr}[T[\hat{V}^\mu,\hat{V}^\nu]]
+g'g^{\prime\prime}\alpha_{34}\epsilon^{\mu\nu\rho\lambda}B_{\mu\nu}X_{\rho\lambda}\nonumber\\
&&+gg^{\prime\prime}\alpha_{35}\epsilon^{\mu\nu\rho\lambda}X_{\mu\nu}\mathrm{tr}[TW_{\rho\lambda}]
+ig'\alpha_{36}\mathrm{tr}[\hat{V}_\mu]\mathrm{tr}[T\hat{V}_\nu]B^{\mu\nu}
+ig\alpha_{37}\mathrm{tr}[\hat{V}_\mu]\mathrm{tr}[T\hat{V}_\nu]\mathrm{tr}[TW^{\mu\nu}]\nonumber\\
&&+g\alpha_{38}\mathrm{tr}[\hat{V}^\mu]\mathrm{tr}[\hat{V}^\nu
W_{\mu\nu}]
+g^{\prime\prime}\alpha_{39}\mathrm{tr}[\hat{V}_\mu]\mathrm{tr}[T\hat{V}_\nu]X^{\mu\nu}
+ig\alpha_{40}\mathrm{tr}[\hat{V}^\mu]\mathrm{tr}[T\hat{V}^\nu
W_{\mu\nu}]\;.
\end{eqnarray}
Among them $\alpha_{12}\sim\alpha_{14}$, $\alpha_{30}$,
$\alpha_{33}\sim \alpha_{40}$ are CP-violation terms. Bosonic part
with differential of higgs field $\mathcal{L}_H$ is
\begin{eqnarray}
\mathcal{L}_H &=&(\partial_\mu
h)\Big\{\alpha_{H,1}\mathrm{tr}[T\hat{V}^\mu]\mathrm{tr}[\hat{V}_\nu\hat{V}^\nu]
    +\alpha_{H,2}\mathrm{tr}[T\hat{V}_\nu]\mathrm{tr}[\hat{V}^\mu\hat{V}^\nu]
    +\alpha_{H,3}\mathrm{tr}[T\hat{V}_\nu]\mathrm{tr}[T[\hat{V}^\mu,\hat{V}^\nu]]
    \nonumber\\
    &&+\alpha_{H,4}\mathrm{tr}[T\hat{V}^\mu]\mathrm{tr}[T\hat{V}_\nu]\mathrm{tr}[T\hat{V}^\nu]
    +ig\alpha_{H,5}\mathrm{tr}[T\hat{V}_\nu]\mathrm{tr}[TW^{\mu\nu}]
    +g'\alpha_{H,6}\mathrm{tr}[T\hat{V}_\nu]B^{\mu\nu}
    \nonumber\\
    &&+ig\alpha_{H,7}\mathrm{tr}[T\hat{V}_\nu W^{\mu\nu}]+g\alpha_{H,8}\mathrm{tr}[\hat{V}_\nu W^{\mu\nu}]
    +\alpha_{H,9}\mathrm{tr}[\hat{V}^\mu]\mathrm{tr}[\hat{V}_\nu\hat{V}^\nu]+\alpha_{H,10}\mathrm{tr}[\hat{V}_\nu]\mathrm{tr}[\hat{V}^\mu\hat{V}^\nu]
    \nonumber\\
    &&+\alpha_{H,11}\mathrm{tr}[\hat{V}_\nu]\mathrm{tr}[T[\hat{V}^\mu,\hat{V}^\nu]]
    +\alpha_{H,12}\mathrm{tr}[\hat{V}^\mu]\mathrm{tr}[T\hat{V}_\nu]\mathrm{tr}[T\hat{V}^\nu]
    +\alpha_{H,13}\mathrm{tr}[\hat{V}^\mu]\mathrm{tr}[\hat{V}_\nu]\mathrm{tr}[T\hat{V}^\nu]\nonumber\\
    &&
    +ig\alpha_{H,14}\mathrm{tr}[\hat{V}_\nu]\mathrm{tr}[TW^{\mu\nu}]
    +g'\alpha_{H,15}\mathrm{tr}[\hat{V}_\nu]B^{\mu\nu}\Big\}
+(\partial_\mu h)(\partial_\nu h)\Big\{
    \alpha_{H,16}\mathrm{tr}[T\hat{V}^\mu]\mathrm{tr}[T\hat{V}^\nu]\nonumber\\
&&+\alpha_{H,17}\mathrm{tr}[\hat{V}^\mu\hat{V}^\nu]
    +\alpha_{H,18}\mathrm{tr}[\hat{V}^\mu]\mathrm{tr}[T\hat{V}^\nu]
    +\alpha_{H,19}\mathrm{tr}[\hat{V}^\mu]\mathrm{tr}[\hat{V}^\nu]\Big\}
+(\partial_\mu h)(\partial^\mu h)\nonumber\\
&&\times\Big\{\alpha_{H,20}\mathrm{tr}[T\hat{V}_\nu]\mathrm{tr}[T\hat{V}^\nu]
    +\alpha_{H,21}\mathrm{tr}[\hat{V}_\nu\hat{V}^\nu]
    +\alpha_{H,22}\mathrm{tr}[\hat{V}_\nu]\mathrm{tr}[T\hat{V}^\nu]
    +\alpha_{H,23}\mathrm{tr}[\hat{V}_\nu]\mathrm{tr}[\hat{V}^\nu]\Big\}
\nonumber\\
&&+(\partial_\mu h)(\partial^\mu h)(\partial_\nu h)\Big\{
    \alpha_{H,24}\mathrm{tr}[T\hat{V}^\nu]
    +\alpha_{H,25}\mathrm{tr}[\hat{V}^\nu]\Big\}
+\alpha_{H,26}[(\partial_\mu h)(\partial^\mu h)]^2\;.
\end{eqnarray}
Anomaly part $\mathcal{L}_A$ is
\begin{eqnarray}
\mathcal{L}_A
&=&\alpha_{42}g^2\epsilon^{\mu\nu\rho\lambda}tr[W_{\mu\nu}W_{\rho\lambda}]
    +\alpha_{43}g'^2\epsilon^{\mu\nu\rho\lambda}B_{\mu\nu}B_{\rho\lambda}
    +{g^{\prime\prime}}^2\alpha_{44}\epsilon^{\mu\nu\rho\lambda}X_{\mu\nu}X_{\rho\lambda}\;.
\end{eqnarray}
Similar as $p^0$ and $p^2$ order, all $\alpha$ coefficients in $p^4$
order Lagrangian are functions of higgs field $h$.

Above chiral Lagrangian is the most general EWCL involve $Z'$ and
higgs fields, in terms of which we can examine details of $Z'$
physics. In following of this section, we focus our attentions on
the mixing among gauge bosons.

We take unitary gauge $\hat{U}=1$. The gauge boson mass term
$\mathcal{L}_M$ and kinetic term $\mathcal{L}_K$ become
\begin{eqnarray}
\mathcal{L}_M &=&\frac{1}{8}f^2g^2\Big[W^1_\mu W^{1,\mu}+W^2_\mu
W^{2,\mu}\Big]
    +\frac{1}{8}(1-2\beta_1)f^2(gW^3_\mu-g'B_\mu)(gW^{3,\mu}-g'B^\mu)\label{LNM}
    \\
    &&+\frac{1}{2}(1-2\beta_3)f^2(g^{\prime\prime}X_\mu+\tilde{g}'B_\mu)(g^{\prime\prime}X^\mu+\tilde{g}'B^\mu)
    +\frac{1}{2}\beta_2f^2(g^{\prime\prime}X_\mu+\tilde{g}'B_\mu)(gW^{3,\mu}-g'B^\mu)\;,\nonumber\\
\mathcal{L}_K&=&
    -\frac{1}{4}B_{\mu\nu}B_{\mu\nu}
    -\frac{1}{4}X_{\mu\nu}X^{\mu\nu}
    -\frac{1}{4}(\partial_\mu W^1_\nu-\partial_\nu W^1_\mu)^2
    -\frac{1}{4}(\partial_\mu W^2_\nu-\partial_\nu W^2_\mu)^2
    \nonumber\\
    &&-\frac{1}{4}(1-\alpha_8g^2)(\partial_\mu W^3_\nu-\partial_\nu W^3_\mu)^2
    +\frac{1}{2}\alpha_1gg'B_{\mu\nu}(\partial_\mu W^3_\nu-\partial_\nu
    W^3_\mu)\nonumber\\
   &&+gg^{\prime\prime}\alpha_{24}X^{\mu\nu}(\partial_\mu W^3_\nu-\partial_\nu W^3_\mu)
    +g'g^{\prime\prime}\alpha_{25}B_{\mu\nu}X^{\mu\nu}\;,\label{LNK}
\end{eqnarray}
in which the charged gauge bosons $W^1_\mu$ and $W^2_\mu$ are
automatically diagonalized. This is due to the fact that there is no
other charged vector and scalar particles to mix with. To generate
mixing for charged gauge bosons, we need to add in theory new
charged gauge bosons, such as $W^{\prime,1}_\mu$ and
$W^{\prime,2}_\mu$ which was already discussed in Ref.\cite{LRSM} or
new charged Higgs bosons. For the remaining neutral gauge bosons
$W^3,B,X$,  our Lagrangian includes most general mixing among them.
We can choose special parameters to recover the various scenarios
discussed in literature. For example,
\begin{itemize}
\item Taking $fg''=M_1$, $f\tilde{g}'=M_2$ and $\alpha_i=\beta_j=0$,
(\ref{LNM}) and (\ref{LNK}) come back to Stueckelberg Lagrangian
given in Ref.\cite{KorsNathinJHEP2005} which depends on coefficients
$M_1$ and $M_2$ .
\item Taking $\frac{\beta_2\sqrt{g^2+g'^2}}{2(1-\beta_3)g''}=x$,
$-2{g''}(g\alpha_{24}-g'\alpha_{25})=y$,
$-2{g''}(g\alpha_{24}+g'\alpha_{25})=w$,
$(1-2\beta_3){g''}^2f^2=m_X^2$ and
$\beta_1=\alpha_i=0~~(i\neq24,25)$, (\ref{LNM}) and (\ref{LNK}) come
back to effective Lagrangian given in Ref.\cite{HoldomSTU1991} which
depends on coefficients $x,y,w$, $m_X^2$ and includes a more
simplified case discussed in an earlier Ref.\cite{HoldomSTU1986}.
\item Taking $-2g{g''}\alpha_{25}=\sin\chi$, and $\beta_1=\beta_3=\alpha_i=0~~~(i\neq 25)$. (\ref{LNM}) and
(\ref{LNK}) come back to effective Lagrangian for $E_6$ model given
in Ref.\cite{Rizzo99} which depends on a mixing angle $\chi$ .
\end{itemize}
What we need to do next is to diagonalize these mass and kinetic
terms. We first try to cancel term
$(g^{\prime\prime}X_\mu+\tilde{g}'B_\mu)(gW^3_\mu-g'B_\mu)$ in
$\mathcal{L}_M$ by mixing $g^{\prime\prime}X+\tilde{g}'B$ with
$gW^3_\mu-g'B_\mu$
\begin{eqnarray}
g^{\prime\prime}X_\mu+\tilde{g}'B_\mu&=&{\rm
cos}\alpha_{Z'}(g^{\prime\prime}\bar{X}_\mu+\tilde{g}'\bar{B}_\mu)+{\rm
sin}\alpha_{Z'}(g\bar{W}^3_\mu-g'\bar{B}_\mu)
\nonumber\\
gW^3_\mu-g'B_\mu&=&-{\rm
sin}\alpha_{Z'}(g^{\prime\prime}\bar{X}_\mu+\tilde{g}'\bar{B}_\mu)+{\rm
cos}\alpha_{Z'}(g\bar{W}^3_\mu-g'\bar{B}_\mu)\;,\label{W3-B}
\end{eqnarray}
where
\begin{eqnarray}
\tan\alpha_{Z'} &=&\frac{3+2\beta_1-8\beta_3-\sqrt{
(3+2\beta_1-8\beta_3)^2 +16\beta_2^2}}{4\beta_2}\label{alphaZ'}.
\end{eqnarray}
$\mathcal{L}_M$ then reads as
\begin{eqnarray}
\mathcal{L}_M =\frac{1}{8}f^2g^2\Big[W^1_\mu W^{1,\mu}+W^2_\mu
W^{2,\mu}\Big]
    +\frac{1}{2}A_2^2f^2
        (g^{\prime\prime}\bar{X}_\mu+\tilde{g}'\bar{B}_\mu)^2
    +\frac{1}{2}A_1^2f^2
        (g\bar{W}^3_\mu-g'\bar{B}_\mu)^2~~\;,
\end{eqnarray}
with
\begin{eqnarray}
A_1^2&=&\frac{1}{4}(1-2\beta_1)c_\alpha^2+\beta_2s_\alpha c_\alpha+(1-2\beta_3)s_\alpha^2
\\
A_2^2&=&\frac{1}{4}(1-2\beta_1)s_\alpha^2-\beta_2s_\alpha
c_\alpha+(1-2\beta_3)c_\alpha^2\;,
\end{eqnarray}
where, $c_\alpha\equiv \cos\alpha_{Z'}$ and
$s_\alpha\equiv\sin\alpha_{Z'}$.

The kinetic term for neutral gauge boson can be written as
\begin{eqnarray}
\mathcal{L}_\mathrm{K,neural}=(W^3_{\mu\nu},B_{\mu\nu},X_{\mu\nu})
    \left(\begin{array}{ccc}
    -\frac{1}{4}(1-\alpha_8g^2)&\frac{1}{4}\alpha_1 gg'&\frac{1}{2}gg^{\prime\prime}\alpha_{24}\\
    \frac{1}{4}\alpha_1 gg'&-\frac{1}{4}&\frac{1}{2}g'g^{\prime\prime}\alpha_{25}\\
    \frac{1}{2}gg^{\prime\prime}\alpha_{24}&\frac{1}{2}g'g^{\prime\prime}\alpha_{25}&-\frac{1}{4}\\
\end{array}\right)\left(\begin{array}{c}W^{3\mu\nu}\\B^{\mu\nu}\\X^{\mu\nu}\end{array}\right)\;.\label{NeutralKenetic0}
\end{eqnarray}
 Decompose $gW^3_\mu$ as
 $(gW^3_\mu-g'B_\mu)/2+(gW^3_\mu+g'B_\mu)/2$,
  $g'B_\mu$ as $-(gW^3_\mu-g'B_\mu)/2+(gW^3_\mu+g'B_\mu)/2$ and $g^{\prime\prime}X_\mu$ as
$g^{\prime\prime}X_\mu+\tilde{g}'B_\mu+(gW^3_\mu-g'B_\mu)\tilde{g}'/2g'-(gW^3_\mu+g'B_\mu)\tilde{g}'/2g'$.
With help of (\ref{W3-B}), we find
\begin{eqnarray}
\left(\begin{array}{c}W^3_\mu\\B_\mu\\X_\mu\end{array}\right)=
\left(\begin{array}{ccc}
    \frac{1}{2g}c_\alpha
        &\frac{1}{2g}
        &-\frac{1}{2g}s_\alpha
    \\
    -\frac{1}{2g'}c_\alpha
        &\frac{1}{2g'}
        &\frac{1}{2g'}s_\alpha
    \\
    \frac{1}{g^{\prime\prime}}(s_\alpha+\frac{\tilde{g}'}{2g'}c_\alpha)
        &-\frac{\tilde{g}'}{2g^{\prime\prime}g'}
        &\frac{1}{g^{\prime\prime}}(c_\alpha-\frac{\tilde{g}'}{2g'}s_\alpha)
    \end{array}\right)
    \left(\begin{array}{c}g\bar{W}^3_\mu-g'\bar{B}_\mu\\
    gW^3_\mu+g'B_\mu\\g^{\prime\prime}\bar{X}_\mu+\tilde{g}'\bar{B}_\mu\end{array}\right)\;.\nonumber
\end{eqnarray}
Further take following transformation which keeps neutral gauge
boson mass terms to be diagonal and rotates neutral gauge boson to
the basis of $Z_\mu$, photon $A_\mu$ and $Z'_\mu$
\begin{eqnarray}
\left(\begin{array}{c}g\bar{W}^3_\mu-g'\bar{B}_\mu\\
gW^3_\mu+g'B_\mu\\g^{\prime\prime}\bar{X}_\mu+\tilde{g}'B_\mu\end{array}\right)
=\left(\begin{array}{ccc}\frac{\cos\beta_{Z'}}{A_1}&0&\frac{\sin\beta_{Z'}}{A_1}\\
ga&gb&gc\\-\frac{\sin\beta_{Z'}}{A_2}&0&\frac{\cos\beta_{Z'}}{A_2}\end{array}\right)
\left(\begin{array}{c}\frac{M_Z}{f}Z_\mu\\A_\mu\\\frac{M_{Z'}}{f}Z'_\mu\end{array}\right)\;.
\end{eqnarray}
 Then the mass term involving neutral gauge bosons can be written as
\begin{eqnarray}
\mathcal{L}_\mathrm{M,neural}&=&\frac{1}{2}A_1^2f^2(g\bar{W}^3_\mu-g'\bar{B}_\mu)+\frac{1}{2}A_1^2f^2\bar{X}_\mu^2
=\frac{1}{2}M_Z^2Z_\mu^2+\frac{1}{2}M_{Z'}^2Z_\mu^{\prime
2}\;,\label{MZMZ'def}
\end{eqnarray}
with massless photon. Remaining six parameters are $Z$ mass $M_Z$,
$Z'$ mass $M_{Z'}$, mixing angle $\beta_{Z'}$ and coefficients
$a,b,c$, which will be determined later. Now total rotation matrix
becomes
\begin{eqnarray}
\left(\begin{array}{c}W^3_\mu\\B_\mu\\X_\mu\end{array}\right)=
U\left(\begin{array}{c}Z_\mu\\A_\mu\\Z'_\mu\end{array}\right)
\end{eqnarray}
with
\begin{eqnarray}
U\equiv\left(\begin{array}{ccc}
    \frac{1}{2g}c_\alpha
        &\frac{1}{2g}
        &-\frac{1}{2g}s_\alpha
    \\
    -\frac{1}{2g'}c_\alpha
        &\frac{1}{2g'}
        &\frac{1}{2g'}s_\alpha
    \\
    \frac{1}{g^{\prime\prime}}(s_\alpha+\frac{\tilde{g}'}{2g'}c_\alpha)
    &-\frac{\tilde{g}'}{2g^{\prime\prime}g'}
        &\frac{1}{g^{\prime\prime}}(c_\alpha-\frac{\tilde{g}'}{2g'}s_\alpha)
\end{array}\right)
\left(\begin{array}{ccc}
    \frac{c_\beta}{A_1}
        &0
        &\frac{s_\beta}{A_1}
    \\
    ga
        &gb
        &gc
    \\
    -\frac{s_\beta}{A_2}
        &0
        &\frac{c_\beta}{A_2}
\end{array}\right)
\left(\begin{array}{ccc}
    \frac{M_Z}{f}&0&0\\
    0&1&0\\
    0&0&\frac{M_{Z'}}{f}
\end{array}
\right)\;, ~~~~~\label{Udef}
\end{eqnarray}
where, $s_\beta=\sin\beta_{Z'}$ and $c_\beta=\cos\beta_{Z'}$. With
above rotation, kinetic term for neutral gauge boson
(\ref{NeutralKenetic0}) can be further written as
\begin{eqnarray}
\mathcal{L}_\mathrm{K,neural}\!=\!(Z_{\mu\nu},A_{\mu\nu},Z'_{\mu\nu})\mathbf{K}\!
\left(\begin{array}{c}Z^{\mu\nu}\\A^{\mu\nu}\\Z^{\prime,\mu\nu}\end{array}\right)\hspace{0.5cm}
\mathbf{K}\!\equiv\! U^T\!\left(\begin{array}{ccc}
    -\frac{1}{4}(1-\alpha_8g^2)&\frac{1}{4}\alpha_1 gg'&\frac{1}{2}gg^{\prime\prime}\alpha_{24}\\
    \frac{1}{4}\alpha_1 gg'&-\frac{1}{4}&\frac{1}{2}g'g^{\prime\prime}\alpha_{25}\\
    \frac{1}{2}gg^{\prime\prime}\alpha_{24}&\frac{1}{2}g'g^{\prime\prime}\alpha_{25}&-\frac{1}{4}\\
\end{array}\right)U\;.~~~~\label{NeutralKenetic}
\end{eqnarray}
In which $\mathbf{K}$ is three by three symmetric matrix. Denote its
matrix elements as $\mathbf{K}_{ij}$. Notice that $\mathbf{K}_{11}
\propto M_Z^2/f^2$ and $\mathbf{K}_{33} \propto M_{Z'}^2/f^2$, then
 normalization of $Z$ and $Z'$ kinetic terms,
\begin{eqnarray}
\mathbf{K}_{11}=-\frac{1}{4}\hspace{2cm}
\mathbf{K}_{33}=-\frac{1}{4}\;,
\end{eqnarray}
is necessary to interpret $M_Z$ and $M_{Z'}$ introduced in
(\ref{MZMZ'def}) as the correct definition of $Z$ and $Z'$ masses.
Above normalization condition also fix values of $M_Z$ and $M_{Z'}$.
Remaining normalization of photon kinetic term demands
$\mathbf{K}_{22}=-1/4$ and diagonalization of kenetic terms requires
$\mathbf{K}_{12}=\mathbf{K}_{13}=\mathbf{K}_{23}=0$. These four
constraint conditions further fix remaining four parameters
$\beta_{Z'},a,b,c$. Detailed computation shows that first
$\mathbf{K}_{22}=-\frac{1}{4}$ fix parameter $b$,
\begin{eqnarray}
b^2=\frac{4{g'}^2{g^{\prime\prime}}^2}{(g^2+{g'}^2){g^{\prime\prime}}^2+g^2\tilde{g}^{\prime2}-{g}^2{g'}^2{g^{\prime\prime}}^2(2\alpha_1+\alpha_8)
    +4g^2g'{g^{\prime\prime}}^2\tilde{g}'(\alpha_{24}+\alpha_{25})}\;.
\nonumber
\end{eqnarray}
 while $\mathbf{K}_{12}=0$ fix parameter $a$,
\begin{eqnarray}
a&=&\frac{1}{gA_1A_2[{g'}^2{g^{\prime\prime}}^2-{g}^2{g'}^2{g^{\prime\prime}}^2(2\alpha_1+\alpha_8)
    +g^2{g^{\prime\prime}}^2
    -4g^2g'{g^{\prime\prime}}^2\tilde{g}'(\alpha_{24}+\alpha_{25})+g^2\tilde{g}^{\prime 2}]}
    \nonumber\\
    &&\times\Big\{[g^2{g''}^2+g^2\tilde{g}^{\prime 2}-g'^2{g''}^2+g^2g'^2{g''}^2\alpha_8+4g^2g'{g''}^2\tilde{g}'\alpha_{25}]
    (s_\alpha s_\beta A_1+c_\alpha c_\beta A_2)\nonumber\\
    &&+[2g^2g'\tilde{g}'+4g^2g'^2{g''}^2(\alpha_{24}+\alpha_{25})](-c_\alpha s_\beta A_1+s_\alpha c_\beta A_2)\Big\}\;. \nonumber
\end{eqnarray}
 $\mathbf{K}_{23}=0$ fix parameter $c$,
\begin{eqnarray}
c&=&\frac{1}{gA_1A_2[{g'}^2{g^{\prime\prime}}^2-{g}^2{g'}^2{g^{\prime\prime}}^2(2\alpha_1+\alpha_8)
    +g^2{g^{\prime\prime}}^2
    -4g^2g'{g^{\prime\prime}}^2\tilde{g}'(\alpha_{24}+\alpha_{25})+g^2\tilde{g}^{\prime2}]}
    \nonumber\\
    &&\times\Big\{[g^2{g''}^2+g^2\tilde{g}^{\prime2}-g'^2{g''}^2+g^2g'^2{g''}^2\alpha_8
    +4g^2g'{g''}^2\tilde{g}'\alpha_{25}](-s_\alpha c_\beta A_1+c_\alpha s_\beta A_2)\nonumber\\
    &&
    +[2g^2g'\tilde{g}'+4g^2g'^2{g''}^2(\alpha_{24}+\alpha_{25})](c_\alpha c_\beta A_1+s_\alpha s_\beta A_2)
    \Big\}\;.~~~~
\nonumber
\end{eqnarray}
 Finally $\mathbf{K}_{1,3}=0$ gives constraint
\begin{eqnarray}
0&=&G_0(1-2c_\beta^2)+G_2s_\beta c_\beta\;, \label{betaequation}
\end{eqnarray}
with $G_0$ and $G_1$ given in (\ref{G2}). Eq.(\ref{betaequation})
yielding $\tan\beta$
\begin{eqnarray}
\tan\beta&=&\frac{-G_2+\sqrt{G_2^2+4G_0^2}}{2G_0}\;.
\end{eqnarray}
 Since the precision
of our computation is only accurate up order of $p^4$, while
$\beta_i,~i=1,2,3,4$ represent $p^2$ order operators and $\alpha_i$
represent $p^4$ order operators. Therefore we can expand our result
in powers of $\beta_i$ and $\alpha_i$ and to our $p^4$ order
precision, we only need to keep terms at most quadratic in $\beta_i$
and linear in $\alpha_i$. Detailed computation gives
$\tan\alpha_{Z'}=-2\beta_2/(3-2\beta_1+8\beta_3)$,
$A_1=(1-\beta_1)/2-\beta_2^2/3-\beta_1^2/4$ and
$A_2=1-\beta_3+\beta_2^2/6-\beta_3^2/2$. The rotation matrix $U$ is
given in (\ref{Uresult}). In literature,  most models
\cite{FranziniPRD1987}\cite{CheyBondZprime}\cite{FramptonPRD1996}\cite{RizzoPRD1991ZprimeMixing}
treat extra $X$ gauge boson by only mixing it with $W^3$ boson in
mass term. This corresponds to
$\alpha_1=\alpha_8=\alpha_{24}=\alpha_{25}=\beta_1=\beta_3=\tilde{g}'=0$
in our EEWCL. The result mixing angle become
\begin{eqnarray}
\tan\xi=\frac{4g^{\prime\prime}\sqrt{g^2+g'^2}}{g^2+g'^2-4{g^{\prime\prime}}^2}\beta_2\;.
\end{eqnarray}
Usually constraints on $\xi$ are highly model-dependent
\cite{PDG06}, the typical value of which is at order of  $10^{-3}$.
In our general case,  $X$ boson can mix not only with $W^3$, but
also with $B$, {\it "There are no quantum numbers which forbid a
mixing of neutral gauge bosons"}\cite{Leike1999}. In Leike's review
article\cite{Leike1999}, general mixing among $X_\mu$, $W^3_\mu$ and
$B_\mu$ in mass terms is parameterized. Further mixing can happen
not only in mass terms, but also in kinetic terms. Authors in
Ref.\cite{HoldomSTU1991}\cite{KeneticMixing} studied a case that in
kinetic terms, there are mixing among $X_\mu$, $W^3_\mu$ and
$B_\mu$. In our formulation, we use three by three $U$ matrix to
parameterize the most general mixing among $X_\mu$, $W^3_\mu$ and
$B_\mu$ happened both in mass terms and kinetic terms. The small
values for mixing among $X$ with $W^3$ and $B$ require smallness in
values for $U_{1,3}, U_{2,3}, U_{3,1}, U_{3,2}$, which from
(\ref{Uresult}) leads to the requirements
\begin{eqnarray}
4{g^{\prime\prime}}^2\neq g^2+g'^2\hspace{1cm}\tilde{g}'\ll
1\hspace{1cm}g^{\prime\prime}\beta_2\ll
1\hspace{1cm}g^{\prime\prime}\alpha_{24}\ll
1\hspace{1cm}g^{\prime\prime}\alpha_{25}\ll
1\;.\label{smallalphabeta}
\end{eqnarray}
Another sector which heavily depends on $W^3$, $B$ and $X$ mixing is
the neutral current. The corresponding Lagrangian is $gW^3_\mu
J^{3,\mu}+g'B_\mu J^\mu_Y+g^{\prime\prime}X_\mu J^\mu_X$, in which
except conventional weak isospin third component current $J^{3,\mu}$
and hypercharge current $J^\mu_Y$, we now have extra hidden current
$J^\mu_X$ couple to $X_\mu$ boson. In terms of physical gauge boson
$Z,A,Z'$, the Lagrangian becomes $eJ_{em}^\mu A_\mu+g_ZJ_Z^\mu
Z_\mu+g^{\prime\prime}J_{Z'}^\mu Z'_\mu$. With help of (\ref{Udef}),
we can read out
\begin{eqnarray}
eJ_{em}^\mu&=&gU_{1,2}J^{3,\mu}+g'U_{2,2}J^\mu_Y+g^{\prime\prime}U_{3,2}J^\mu_X\nonumber\\
g_ZJ^\mu_Z&=&gU_{1,1}J^{3,\mu}+g'U_{2,1}J^\mu_Y+g^{\prime\prime}U_{3,1}J^\mu_X\\
g^{\prime\prime}J_{Z'}^\mu&=&gU_{1,3}J^{3,\mu}+g'U_{2,3}J^\mu_Y+g^{\prime\prime}U_{3,3}J^\mu_X\;.\nonumber
\end{eqnarray}
For which, we find
\begin{itemize}
\item When $\tilde{g}'\neq 0$, due to fact $U_{3,2}\neq 0$  given in (\ref{Uresult}), photon will
couple to hidden neutral current $J^\mu_X$. This situation was
discussed in Ref.\cite{KorsNathPLB}.
\item Small mixing among $X$ with $W^3$ and $B$ achieved by
(\ref{smallalphabeta}) will imply that hidden neutral current
$J^\mu_X$ decouples from $Z$ boson and photon approximately;
$J^{3,\mu}$ and $J^\mu_Y$ also decouple from $Z'$ boson
approximately.
\item  $J^\mu_X$ mainly couples to $Z'$ and the coupling
is $g^{\prime\prime}$ which will see later that is proportional to
$M_{Z'}/f$.
\end{itemize}

We now display the last three parameters accurate up order of $p^4$
and linear order of $\tilde{g}'$
\begin{eqnarray}
\beta_{Z'}
&=&\frac{1}{{\Delta_g}}\Big\{2g'\tilde{g}'-\frac{1}{3}(3g_Z^2-{\Delta_g})\beta_2
        -4{g^{\prime\prime}}^2(g^2\alpha_{24}-g'^2\alpha_{25})\Big\}
    -\frac{2}{9{\Delta_g}^2}\Big\{-9g'(2g_Z^2-{\Delta_g})\tilde{g}'(\beta_1-\beta_3)
    \nonumber\\
    &&+2(g_Z^2-{g^{\prime\prime}}^2)\beta_2[(g_Z^2+20{g^{\prime\prime}}^2)\beta_1+(5g_Z^2-44{g^{\prime\prime}}^2)\beta_3]\Big\}
    -\,{\frac {2{g}^{2}\tilde{g}'g' \left({\Delta_g}-2{g'}^{2}\right) \alpha_{{1}}}{ {\Delta_g}^{2}}}
    \nonumber\\
    &&-\,{\frac {2g'\tilde{g}'{g}^{4}\alpha_{{8}}}{{\Delta_g}^{2}}}
    -\,{\frac {2g'\tilde{g}' \left(g_Z^4+24g_Z^2{g''}^{2}+16\,{g''}^{4} \right) \beta_{{3}}\beta_{{1}}}{{\Delta_g}^{3}}}
    +{\frac {g' \left(3g_Z^4+24g_Z^2{g''}^{2}-16\,{g''}^{4}\right) \tilde{g}'{\beta_{{1}}}^{2}}{{\Delta_g}^{3}}}
    \nonumber\\
    &&+\frac{1}{3}\,{\frac {\tilde{g}' \left( 29g_Z^4-16\,{g''}^{4}-136g_Z^2{g''}^{2} \right) g'{\beta_{{2}}}^{2}}{{\Delta_g}^{3}}}
    -{\frac {g' \left( -24g_Z^2{g''}^{2}-48\,{g''}^{4}+g_Z^4\right)\tilde{g}' {\beta_{{3}}}^{2}}{{\Delta_g}^{3}}}
\nonumber\\
\frac{M_Z^2}{f^2}&=&\frac{1}{4}\Big[g_Z^2(1-2\beta_1)-2g^2g'^2\alpha_1+g^4\alpha_8\Big]
    -{\frac { g_Z^2 (g'\tilde{g}'\beta_{{2}}-{g''}^2\beta_2^2)}{{\Delta_g}}}
    \nonumber\\
    &&-\,{\frac {4\tilde{g}'g'{g''}^{2}({g}^{2}\alpha_{{24}}-g'^2\alpha_{25})}{{\Delta_g}}}
    -\,{\frac {8g_Z^2{g''}^{2}g'\tilde{g}'}{{\Delta_g}^{2}}}
        (\beta_{{2}}\beta_{{1}}-\beta_2\beta_3)
\nonumber\\
\frac{M_{Z'}^2}{f^2}&=&{g^{\prime\prime}}^2(1-2\beta_3)
    +\frac{{g^{\prime\prime}}^2[4g'\tilde{g}'\beta_2-g_Z^2\beta_2^2]}{{\Delta_g}}
    +4\,{\frac {{g''}^{2}\tilde{g}'g'[{g}^{2}\alpha_{{24}}-(g^2-4{g''}^2)   \alpha_{{25}}]}{{\Delta_g}}}
    \nonumber\\
    &&+8\,{\frac {{g''}^{2}g'\tilde{g}'g_Z^2}{{\Delta_g}^{2}}}
        (\beta_{{2}}\beta_{{1}}-\beta_2\beta_3)\;,
\end{eqnarray}
where $g_Z=\sqrt{g^2+g'^2}$ and ${\Delta_g}=g^2+g'^2-4{g''}^2$. All
$\beta_i$ and $\alpha_i$ coefficients appear in above results must
take their values with Higgs field inside the coefficients being
substituted by its vacuum expectation value. Notice that the
correction for $Z$ mass from extra $Z'$ couplings is proportional to
$\tilde{g}'\beta_2$, $(g^{\prime\prime}\beta_2)^2$, $g''\alpha_{24}$
and $g''\alpha_{25}$ which are very small if we adopt
(\ref{smallalphabeta}). In fact, in formulae for $M_Z$ and $M_{Z'}$,
if we ignore these small mixing and further neglect contribution
from $\beta_1,\alpha_1,\alpha_8$ which roughly are related to
phenomenological parameters $T,S,U$\cite{Appelquist93}, we find
$M_{Z'}^2/M_Z\sim{2g^{\prime\prime}}^2(1-2\beta_3)/e^2$. This
implies that even for small mixing for neutral gauge bosons with
$Z'$ we still have two independent parameters $g^{\prime\prime}$ and
$\beta_3$ to tune its value.

We finish discussion on mixing among neutral gauge bosons by
checking our computation results. With constraints
(\ref{smallalphabeta}), $X$ mixing with $W^3$ and $B$ controlled by
parameter $\tilde{g}'$, $g^{\prime\prime}\beta_2$,
$g^{\prime\prime}\alpha_{24}$ and $g^{\prime\prime}\alpha_{25}$
become very small. Ignoring contributions from these parameters, $X$
will not mix with $W^3$ and $B$ any more and the left mixing between
$W^3$ and $B$ then goes back to its value given be standard
EWCL\cite{Appelquist93}. If we further demand
$\tilde{g}'=\beta_1=\beta_2=\beta_3=\alpha_1=\alpha_8=\alpha_{24}=\alpha_{25}=0$,
we recover results of tree diagram SM which include
$G_0=0,~A_1=1/2,~A_2=1$ and the matrix $U$ becomes
\begin{eqnarray}
    \left(\begin{array}{ccc}
    \cos\theta_W & \sin\theta_W & 0 \\
    -\sin\theta_W & \cos\theta_W & 0 \\
    0 & 0 & 1
    \end{array}\right),
\end{eqnarray}
with $\tan\theta_W=g'/g$. The six parameters at this order of
approximation are $\alpha_{Z'}=\beta_{Z'}=c=0$,
$a=2\frac{g^2-g'^2}{g(g^2+g'^2)}$,
$b=\frac{2g'}{\sqrt{g^2+{g'}^2}}$, $M_Z=\frac{f}{2}\sqrt{g^2+g'^2}$
and $M_{Z'}={g''}f$.

%%%%%%%%%%%%%%%%%%%%%%%%%%%%%%%%%%%%%%%%
\section{Anomalous couplings among gauge fields}

In this section, we discuss effective gauge boson self interactions
which include  triple and quartic coupling terms and these terms
which not only include SM electroweak gauge fields $W^\pm,A,Z$, but
also involve $Z'$ field. The part without $Z'$ field can be
parameterized by coefficients in original EWCL and parametrization
for quadratic and triple couplings were already given in
Ref.\cite{Appelquist93}. The quartic couplings can be worked out as
follows,
\begin{eqnarray}
\mathcal{L}_\mathrm{QGV}&=&g_{++--}W^+_{\mu}W^{+\mu}W^-_{\nu}W^{-\nu}-g_{+-+-}
(W^+_{\mu}W^{-\mu})^2 +g_{Z^4}(Z_{\mu}Z^{\mu})^2+g_{+Z-Z}W^+_{\mu}Z^{\mu}W^-_{\nu}Z^{\nu}\nonumber\\
&&-g_{+-ZZ}W^+_{\mu}W^{-\mu}Z_{\nu}Z^{\nu}-g_{+-ZA}W^+_{\mu}W^{-\mu}Z_{\nu}A^{\nu}-ig_{\epsilon+-ZA}\epsilon^{\mu\nu\rho\lambda}
Z_{\mu}A_{\nu}W^+_{\rho}W^-_{\lambda}\nonumber\\
&&+[g_{+Z-A\oplus+A-Z}(W^+_{\mu}W^-_{\nu}+W^+_{\nu}W^-_{\mu})
+ig_{+Z-A\ominus+A-Z}(W^+_{\mu}W^-_{\nu}-W^+_{\nu}W^-_{\mu})]Z^{\mu}A^{\nu}\nonumber\\
&&-e^{*2}(A_{\mu}A^{\mu}W^{+\nu}W^-_{\nu}-W^-_{\mu}W^+_{\nu}A^{\mu}A^\nu)\;,\nonumber
\end{eqnarray}
with nine anomalous quartic couplings determined by
\begin{eqnarray}
&&\hspace{-0.5cm}g_{++--}=\frac{e^{*2}}{8\sin^2\theta_W}|_Z[4
+\frac{2\alpha_1e^2}{c^2\!-\!s^2}+\alpha_8e^2(-\frac{1}{s^2}\!+\!\frac{c^2}{c^2\!-\!s^2})\!+\!\frac{2\beta_1c^2}{c^2\!-\!s^2}
+(\alpha_3\!\!+\!\frac{1}{2}\alpha_4\!\!-\!\frac{1}{2}\alpha_8\!+\!\alpha_9)\frac{8e^2}{s^2}]\nonumber\\
&&\hspace{-0.5cm}g_{+-+-}=-\frac{e^{*2}}{8\sin^2\theta_W}|_Z
[4+\frac{2\alpha_1e^2}{c^2\!-\!s^2}+\alpha_8e^2(-\frac{1}{s^2}\!+\!\frac{c^2}{c^2\!-\!s^2})
+\frac{2\beta_1c^2}{c^2\!-\!s^2}
-(-\alpha_3\!\!+\!\frac{1}{2}\alpha_4\!\!+\!\alpha_5\!\!+\!\frac{1}{2}\alpha_8\!-\!\alpha_9)\frac{8e^2}{s^2}]\nonumber\\
&&\hspace{-0.5cm}g_{Z^4}=e^{*4}\cot^4\theta_W|_Z
(\frac{1}{2}\alpha_4+\frac{1}{2}\alpha_5+\alpha_6+\alpha_7+\alpha_{10})\frac{1}{2c^8}\nonumber\\
&&\hspace{-0.5cm}g_{+Z-Z}=e^{*2}\cot^2\theta_W|_Z[1+
\frac{2\beta_1}{c^2-s^2}+\frac{2}{c^2(c^2-s^2)}e^2\alpha_1
+(2\alpha_3+\frac{\alpha_4\!+\!\alpha_6}{c^2})\frac{e^2}{s^2c^2}]\nonumber\\
&&\hspace{-0.5cm}g_{+-ZZ}=e^{*2}\cot^2\theta_W|_Z[1
+\frac{2\beta_1}{c^2-s^2}+\frac{2}{c^2(c^2-s^2)}e^2\alpha_1
+(2\alpha_3-\frac{\alpha_5+\alpha_7}{c^2})\frac{e^2}{s^2c^2}]\nonumber\\
&&\hspace{-0.5cm}g_{+-ZA}=2[1+\frac{\beta_1}{c^2-s^2}+\frac{e^2\alpha_1}{c^2(c^2-s^2)}
+\alpha_3\frac{e^2}{s^2c^2}]\hspace{1cm}g_{\epsilon+-ZA}=\frac{2e^2}{s^2c^2}\alpha_{11}\\
&&\hspace{-0.5cm}g_{+Z-A\oplus+A-Z}=e^{*2}\!\cot\theta_W|_Z[1+
\frac{\beta_1}{c^2\!-\!s^2}+\frac{1}{c^2(c^2\!-\!s^2)}e^2\alpha_1
+\alpha_3\frac{e^2}{s^2c^2}]\hspace{0.7cm}g_{+Z-A\ominus+A-Z}=\alpha_{12}\frac{e^2}{s^2c^2}\;,\nonumber
\end{eqnarray}
where all coefficients are defined in Ref.\cite{Appelquist93}. We
can also obtain these anomalous couplings from our theory by taking
unitary gauge $\hat{U}=1$. Matching these anomalous couplings from
original EWCL with those obtained from our theory involving $Z'$
boson, we obtain constraints which relate parameters in original
EWCL with those in ours. These constraints can be seen as an
alternative result obtained through integrating out $Z'$ field and
its goldstone boson. Some of them are not independent each other and
can be treated as self consistent check of our computation. Detailed
matching for $M^2_{W^\pm}$ demands that the fundamental parameter
$f$ in (\ref{L2}) be the same as that introduced in original EWCL
\cite{Appelquist93}. Matching for $M^2_Z$ gives
\begin{eqnarray}
\delta\beta_1 &=&g^2\alpha_3
    +\frac{2(g^2+{\Delta_g})}{{\Delta_g}^2}(g'\tilde{g}'\beta_2-{g''}^2\beta_2^2)
    +\frac{2g^4g'\tilde{g}'}{{\Delta_g}^2}\alpha_{24}
    -\frac{2g'\tilde{g}'[4g'^2{g''}^2-(g^2-4{g''}^2)^2]}{{\Delta_g}^2}\alpha_{25}
    \nonumber\\
    &&+\frac{4g'\tilde{g}'}{{\Delta_g}^3}
        [(4{g''}^2{\Delta_g}-g^2{\Delta_g}+2g^2g_Z^2)\beta_1\beta_2-4{g''}^2(2g^2+{\Delta_g})\beta_2\beta_3]\;,\label{deltabeta}
\end{eqnarray}
with $\delta\beta_1$ being the difference between $\beta_1$
introduced in (\ref{L2}) $\beta_1\Big|_{Z'}$ and corresponding
parameter introduced in original EWCL $\beta_1\Big|_{EWCL}$, i.e.
$\delta\beta_1=\beta_1\Big|_{Z'}-\beta_1\Big|_{EWCL}$. While
matching triple and quartic anomalous couplings gives
\begin{eqnarray}
 \delta\alpha_1
&=&\delta\alpha_3+\frac{2}{{\Delta_g}^2}(g'\tilde{g}'\beta_2-{g''}^2\beta_2^2)
    +\frac{4g'\tilde{g}'}{{\Delta_g}^3}[(2g_Z^2-{\Delta_g})\beta_1\beta_2-8{g''}^2\beta_2\beta_3]
    \nonumber\\
    &&+\frac{2\tilde{g}'}{g'{\Delta_g}^2}\Big\{[(g^2-4{g''}^2)^2+2{g''}^2(g^2-2g'^2)]\alpha_{24}
        +(g^2-4{g''}^2)(g'^2+{\Delta_g})\alpha_{25}\Big\}
\nonumber\\
\delta\alpha_2
&=&\frac{2}{{\Delta_g}^2}(g'\tilde{g}'\beta_2-{g''}^2\beta_2^2)
    +\frac{4g'\tilde{g}'}{{\Delta_g}^3}[(2g_Z^2-{\Delta_g})\beta_1\beta_2-8{g''}^2\beta_2\beta_3]
    \nonumber\\
    &&+\frac{2\tilde{g}'}{g'{\Delta_g}^2}\Big\{g^2g'^2\alpha_{24}
        +(g^2-4{g''}^2)(g'^2+{\Delta_g})\alpha_{25}\Big\}
\nonumber\\
\delta\alpha_4
&=&\frac{4}{{\Delta_g}^2}(g'\tilde{g}'\beta_2-{g''}^2\beta_2^2)
    +\frac{8g'\tilde{g}'}{{\Delta_g}^3}[(2g_Z^2-{\Delta_g})\beta_1\beta_2-8{g''}^2\beta_2\beta_3]
    +\frac{4g'\tilde{g}'}{{\Delta_g}^2}(g^2\alpha_{24}-g'^2\alpha_{25})\nonumber\\
   &&-\frac{4g'\tilde{g}'}{{\Delta_g}}\alpha_{31}
\nonumber\\
\delta\alpha_5
&=&-\delta\alpha_4
\nonumber\\
\delta\alpha_6
&=&-\delta\alpha_4-\frac{2g'\tilde{g}'}{{\Delta_g}}\alpha_{17}
\nonumber\\
\delta\alpha_7
&=&\delta
\alpha_4-\frac{2g'\tilde{g}'}{{\Delta_g}}\alpha_{16}
\nonumber\\
\delta\alpha_8
&=&-2\delta\alpha_3
    -\frac{4}{{\Delta_g}^2}(g'\tilde{g}'\beta_2-{g''}^2\beta_2^2)
    -\frac{8g'\tilde{g}'}{{\Delta_g}^3}[(2g_Z^2-{\Delta_g})\beta_1\beta_2-8{g''}^2\beta_2\beta_3]
    \nonumber\\
    &&+\frac{4g'\tilde{g}'}{{\Delta_g}^2}[(g'^2-4{g''}^2)\alpha_{24}+g'^2\alpha_{25}]
\nonumber\\
\delta\alpha_9
&=&-2\delta
\alpha_3
    -\frac{4}{{\Delta_g}^2}(g'\tilde{g}'\beta_2-{g''}^2\beta_2^2)
    -\frac{8g'\tilde{g}'}{{\Delta_g}^3}[(2g_Z^2-{\Delta_g})\beta_1\beta_2-8{g''}^2\beta_2\beta_3]
    \nonumber\\
    &&+\frac{2g'\tilde{g}'}{{\Delta_g}^2}[({\Delta_g}-2g^2)\alpha_{24}+2g'^2\alpha_{25}]
    +\frac{2g'\tilde{g}'}{{\Delta_g}}\alpha_{31}
\nonumber\\
\delta\alpha_{10} &=&-\frac{g'\tilde{g}'}{{\Delta_g}}\alpha_{15}\;,
\label{deltaalphai}
\end{eqnarray}
with $\delta\alpha_i=\alpha_i\Big|_{Z'}-\alpha_i\Big|_{EWCL}$ and
left $\delta\alpha_3$ undetermined. In obtaining above result, we
are accurate up to linear order of $\tilde{g}'$ and neglect all CP
violation coefficients.

Beyond the self interaction part without $Z'$ field, there is part
depending on $Z'$ field. The quadratic term is already discussed
before and we list down the triple and quartic vertices,
\begin{eqnarray}
\mathcal{L}_{Z'~\mathrm{anomalous}}
&=&iC_{Z'-+}{Z'}^{\mu\nu}W^+_\mu W^-_\nu+iC_{+Z'-}(W^+_{\mu\nu}W^{-,\mu}Z'^\nu -W^-_{\mu\nu} W^{+,\mu}Z'^\nu)
    \label{Z'anomalous}\\
    &&+D_{+-V_1V_2}W^+_\mu W^{-,\mu}V_{1,\nu} {V_2}^\nu
    +D_{+V_1-V_2}W^+_\mu V_1^\mu W^-_\nu V_2^\nu
    +D_{V_1V_2V_3V_4}V_{1,\mu}V_2^\mu V_{3,\nu}V_4^\nu\;.
\nonumber
\end{eqnarray}
The explicit expressions for various couplings in above Lagrangian
are given in (\ref{Z'couplings}).

%%%%%%%%%%%%%%%%%%%%%%%%%%%%%%%%%%%%%%%%%%%%%%%%%%%
\section{Summary}

Stueckelberg mechanism as a traditional method to introduce a $U(1)$
gauge boson into theory is shown in this paper equivalent to set up
a gauged $U(1)$ chiral Lagrangian and fix special gauge. With this
equivalence to chiral Lagrangian, by constructing the non-abelian
generalization of the chiral Lagrangian, it is easy to understand
why non-abelian generalization of the Stueckelberg mechanism can not
keep renormalizability. Further in terms of chiral Lagrangian
formulation, we generalize traditional Stueckelberg mechanism by
including in theory high dimension operators. We enlarge original
EEWCL to include an extra local $U(1)$ symmetry to represent physics
for $Z'$ boson. The scalar particle in Stueckelberg mechanism now is
identified with goldstone boson eaten out by $Z'$ to become its
longitudinal component.  We build up complete list of EEWCL up to
order of $p^4$ including $Z'$ and higgs bosons. With this chiral
Lagrangian, traditional minimal version of the Stueckelberg
mechanism can be seen as the leading nonlinear $\sigma$ model term
of our theory and our generalization for Stueckelberg mechanism is
to include in theory all possible high dimension operators up to
order of $p^4$. We obtain most general interaction for$Z'$ boson and
SM bosons. Among these interactions, we focus on the general mixing
among neutral gauge boson $W^3$, $B$ and $X$. We diagonalize the
mixing appeared in mass and kinetic terms completely by introducing
a three by three matrix $U$.  The small mixing among $X$ with $W^3$
and $B$ can be achieved by constraints (\ref{smallalphabeta}). Due
to lack of enough theoretical constraints and experiment data, most
of operators lead by our extension of Stueckelberg mechanism have
their free couplings. We need to gather more theoretical arguments
and experiment data to investigate them in future. Theoretically,
through matching anomalous couplings between original EWCL and our
theory, we obtain connections among parameters in
Ref.\cite{Appelquist93} and those in our theory which enable us to
express anomalous couplings in terms of parameters appeared in our
theory. We also exhibit all $p^4$ order operators for gauge fields
self-interaction involving $Z'$.

%%%%%%%%%%%%%%%%%%%%%%%%%%%%%%%%%%%%%%%%%%%%%
\section*{Acknowledgment}
This work was  supported by National  Science Foundation of China
(NSFC) under Grant No. 10435040 and Specialized Research Fund for
the Doctoral Program of Higher Education.
%%%%%%%%%%%%%%%%%%%%%%%%%%%%%%%%%%%%%%%%%%%%%%%%%%%%
\appendix
\section{Necessary formulae for EWCL}

In this appendix, we list down the necessary lengthy formulae needed
in the text. First we give expression for $G_0$ and $G_1$ introduced
in (\ref{betaequation}),
\begin{eqnarray}
G_0&=&-A_1A_2\Big\{(-g^2-g'^2+{g^{\prime\prime}}^2+(\tilde{g}')^2)c_\alpha
s_\alpha
    +g'\tilde{g}'(s_\alpha^2-c_\alpha^2)
    +g^2[2g'^2c_\alpha s_\alpha+g'\tilde{g}'(c_\alpha^2-s_\alpha^2)]\alpha_1
    \nonumber\\
    &&+g^2[(g'^2-{g^{\prime\prime}}^2-(\tilde{g}')^2)c_\alpha s_\alpha -g'\tilde{g}'(s_\alpha^2-c_\alpha^2)]\alpha_8
    +2g^2{g^{\prime\prime}}^2(c_\alpha^2-s_\alpha^2)(\alpha_{24}+g'^2\alpha_1\alpha_{25})
    \nonumber\\
    &&+{g^{\prime\prime}}^2[-4g'\tilde{g}'c_\alpha s_\alpha+2g'^2(c_\alpha^2-s_\alpha^2)]
        [g^2(\alpha_8\alpha_{25}-\alpha_1\alpha_{24})-\alpha_{25}]+g^2{g^{\prime\prime}}^2[8g'^2s_\alpha c_\alpha
    \nonumber\\
    &&+4g'\tilde{g}'(c_\alpha^2-s_\alpha^2)]\alpha_{24}\alpha_{25}
    +g^2g'^2{g^{\prime\prime}}^2s_\alpha c_\alpha(4\alpha_{25}^2-\alpha_1^2)
   +4g^2{g^{\prime\prime}}^2(g's_\alpha+\tilde{g}'c_\alpha)(g'c_\alpha-\tilde{g}'s_\alpha)\alpha_{24}^2
    \Big\}
\nonumber\\
G_2&=&A_1^2\Big\{(g^2+g'^2)c_\alpha^2
    +({g^{\prime\prime}}^2+(\tilde{g}')^2)s_\alpha^2(1-g^2\alpha_8)
    -g^2g'^2c_\alpha^2(2\alpha_1+\alpha_8)
    +4g'{g^{\prime\prime}}^2\tilde{g}'s_\alpha^2\alpha_{25}
    \nonumber\\
    &&-4g^2g'^2{g^{\prime\prime}}^2c_\alpha^2(\alpha_{24}^2+\alpha_{25}^2+2\alpha_{24}\alpha_{25})
    -g^2{g^{\prime\prime}}^2s_\alpha^2[g'^2\alpha_1^2+4(\tilde{g}')^2\alpha_{24}^2+4g'\tilde{g}'(\alpha_8\alpha_{25}-\alpha_1\alpha_{24})]
    \Big\}
    \nonumber\\
    &&-[A_1\rightarrow A_2,c_\alpha\leftrightarrow s_\alpha]
    +s_\alpha c_\alpha(A_1^2+A_2^2)\Big\{-2g'\tilde{g}'[1-g^2(\alpha_1+\alpha_8)]
    \nonumber\\
    &&+4g^2{g^{\prime\prime}}^2[(\alpha_{24}-\alpha_{25})(1-{g''}^2\alpha_1)+2g'\tilde{g}'\alpha_{24}^2+{g''}^2\alpha_8\alpha_{25}]\Big\}\;.
    \label{G2}
\end{eqnarray}

Next result is for rotation matrix $U$ defined in (\ref{Udef}), its
matrix elements $U_{i,j}$ are
\begin{eqnarray}
U_{1,1}
&=&\frac{g}{g_Z}[1-\frac{g'^4}{g_Z^2}\alpha_1+\frac{g^2(g^2+2g'^2)}{2g_Z^2}\alpha_8]
    +\frac{2gg_Z}{{\Delta_g}^2}(g'\tilde{g}'\beta_2-{g''}^{2}\beta_2^2)
    -8\,{\frac {{g''}^{2}\tilde{g}'g' (g'^2{\Delta_g}-g^2g_Z^2) g\alpha_{{24}}}{{\Delta_g}^2g_Z^3}}
    \nonumber\\
    &&-8\,{\frac {{g'}^{3}\tilde{g}'{g''}^{2} (g_Z^2+{\Delta_g}) g\alpha_{{25}}}{{\Delta_g}^2g_Z^3}}
    +\,{\frac {4\tilde{g}'g'g_Zg}{{\Delta_g}^{3}}}[4{g''}^{2}\beta_2(\beta_{{1}}-2\beta_3)+g_Z^2\beta_{{1}}\beta_{{2}}]
\nonumber\\
U_{1,2} &=&\frac{g'}{g}U_{2,2}
=\frac{g'}{g_Z}[1+\frac{g^2g'^2}{g_Z^2}(\alpha_1+\frac{1}{2}\alpha_8)]
    -\frac{2g^2g'^2\tilde{g}'}{g_Z^3}(\alpha_{24}+\alpha_{25})
\nonumber\\
U_{1,3} &=&\frac{2gg''}{\Delta_g}
        [\frac{g'\tilde{g}'}{2{g''}^2}-\beta_2+(g'^2-4{g''}^2)\alpha_{24}+g'^2\alpha_{25}
        +4g'\tilde{g}'(\beta_1-\beta_3)-2g_Z^2\beta_1\beta_2+8{g''}^2\beta_2\beta_3]
    \nonumber\\
    &&-\frac{2gg'\tilde{g}'(2{g''}^2{\Delta_g}-g^2g'^2)}{g''{\Delta_g}^2}\alpha_1
    +\frac{g^3g'\tilde{g}'(g'^2-4{g''}^2)}{g''{\Delta_g}^2}\alpha_8
    +\frac{16gg'g''\tilde{g}'}{{\Delta_g}^3}
        [g_Z^2\beta_1^2-{\Delta_g}\beta_1\beta_3+4{g''}^2\beta_3^2]
    \nonumber\\
    &&+\frac{2}{3}\frac{gg'g''\tilde{g}'(-112g_Z^2{g''}^2+17g_Z^4+32{g''}^4)}{g_Z^2{\Delta_g}^3}\beta_2^2
\nonumber\\
U_{2,1}
&=&-\frac{g'}{g_Z}[1-\frac{g^4}{g_Z^2}(\alpha_1+\frac{1}{2}\alpha_8)]
    +\frac{2\tilde{g}'(g^2-4{g''}^2)g_Z}{{\Delta_g}^2}\beta_2
    -4\,{\frac {\tilde{g}'g_Z(-g^2{\Delta_g}+8\,{g'}^{2}{g''}^{2}) \beta_{{1}}\beta_{{2}}}{{\Delta_g}^{3}}}
    \nonumber\\
    &&+\frac{8{g''}^2\tilde{g}'}{g_Z^3{\Delta_g}^2}
        [g^2(g^2{\Delta_g}-g'^2g_Z^2)\alpha_{24}-2g'^2(g^2{\Delta_g}-2g'^2{g''}^2)\alpha_{25}]
    +\,{\frac {2{g''}^{2}g'g_Z{\beta_{{2}}}^{2}}{{\Delta_g}^{2}}}
    \nonumber\\
    &&-16\,{\frac {{g''}^{2}\tilde{g}'g_Z({\Delta_g}-2{g'}^{2} ) \beta_{{2}}\beta_{{3}}}{{\Delta_g}^{3}}}
\nonumber\\
U_{2,3} &=&\frac{2g'g''}{{\Delta_g}}
        (\frac{g^2-4{g''}^2}{2g'{g''}^2}\tilde{g}'+\beta_2+g^2\alpha_{24}+(g^2-4{g''}^2)\alpha_{25})
    +\frac{g^2g'^2\tilde{g}'}{g''{\Delta_g}^2}
        [2(g^2-4{g''}^2)\alpha_1+g^2\alpha_8]
    \nonumber\\
    &&+\frac{4g'g''}{{\Delta_g}^2}
        (-2g'\tilde{g}'(\beta_1-\beta_3)+g_Z^2\beta_1\beta_2-4{g''}^2\beta_2\beta_3)
    +\,{\frac {4g''g'}{{\Delta_g}^2}}
         [g_Z^2\beta_{{1}}\beta_{{2}}-4{g''}^2\beta_2\beta_3]
    \nonumber\\
    &&-\frac{2}{3}\frac{g''\tilde{g}'(20g'^6-12g^4{g''}^2-136g^2g'^2{g''}^2-124g'^4{g''}^2+32g'^2{g''}^4+43g^2g'^4+26g^4g'^2+3g^6)}{g_Z^2{\Delta_g}^3}\beta_2^2
    \nonumber\\
    &&-\frac{16g'^2g''\tilde{g}'g_Z^2}{{\Delta_g}^2}\beta_1^2
    +\frac{16g'^2g''\tilde{g}'}{{\Delta_g}^3}
        [(2g_Z^2-{\Delta_g})\beta_1\beta_3-4{g''}^2\beta_3^2]
\nonumber\\
U_{3,1}
&=&\frac{2g''g_Z}{{\Delta_g}}(-\frac{2g'\tilde{g}'}{g_Z^2}+\beta_2+g^2\alpha_{24}-g'^2\alpha_{25})
    +\frac{2g^2g'g''\tilde{g}'}{g_Z^3{\Delta_g}^2}[2(g^2{\Delta_g}-2g_Z^2g'^2)\alpha_1+g^2(2g_Z^2+{\Delta_g})\alpha_8]
    \nonumber\\
    &&+\frac{4g''g_Z}{{\Delta_g}^2}(-2g'\tilde{g}'(\beta_1-\beta_3)
        +g_Z^2\beta_1\beta_2
        -4{g''}^2\beta_2\beta_3)
    -\frac{8}{3}\,{\frac {g'\tilde{g}'g_Z(2g_Z^4-13\,{g''}^{4}-7g_Z^2{g''}^{2} ) {\beta_{{2}}}^{2}}{g''{\Delta_g}^{3}}}
    \nonumber\\
    &&-\frac{16g'g''\tilde{g}'g_Z}{{\Delta_g}^3}[g_Z^2\beta_1^2+4{g''}^2\beta_3^2+({\Delta_g}-2g_Z^2)\beta_1\beta_3]
\nonumber\\
U_{3,2}
&=&-\frac{g\tilde{g}'}{\sqrt{g_Z^2{g''}^2+g^2(\tilde{g}')^2-g^2g'^2{g''}^2(2\alpha_1+\alpha_8)+4g^2g'{g''}^3\tilde{g}'(\alpha_{24}+\alpha_{25})}}
\nonumber\\
&=&-\frac{g\tilde{g}'}{g''g_Z}
    -\frac{g^3g'^2\tilde{g}'}{g''g_Z^3}(\alpha_1+\frac{1}{2}\alpha_8)
\nonumber\\
U_{3,3}
&=&1+\frac{2{g''}^2}{{\Delta_g}^2}[4g'\tilde{g}'(\beta_2+2\beta_2\beta_3+g^2\alpha_{24}-g'^2\alpha_{25})-g_Z^2\beta_2^2]
    +\frac{32g'{g''}^2\tilde{g}'g_Z^2}{{\Delta_g}^3}\beta_2(\beta_1-\beta_3)\;,
\label{Uresult}
\end{eqnarray}
where $g_Z=\sqrt{g^2+g'^2}$ and ${\Delta_g}=g^2+g'^2-4{g''}^2$ and
except $U_{3,2}$ which vanishes when $\tilde{g}'=0$, all other
matrix elements are accurate up to linear order of $\tilde{g}'$.

The last formulae are the anomalous  triple and quartic couplings
for $Z'$ field introduced in (\ref{Z'anomalous}),
\begin{eqnarray}
C_{Z'-+} &=&\,{\frac
{2{g}^{2}g''(\beta_{{2}}+g^2\alpha_{24}-g'^2\alpha_{25})}{{\Delta_g}}}
    +\,{\frac {4{g}^{2}g''[g_Z^2\beta_{{1}}\beta_{{2}}-4{g''}^2\beta_2\beta_3]}{{\Delta_g}^{2}}}
    -2\,{g}^{2}g''\alpha_{{31}}
    \nonumber\\
    &&-\frac{2}{3}\,{\frac {{g}^{2}\tilde{g}'g'g'' ( 17g_Z^{4}-112g_Z^2{g''}^{2}+32\,{g''}^{4}) {\beta_{{2}}}^{2}}{{\Delta_g}^{3}g_Z^2}}
    +{\frac {{g}^{4}g'\tilde{g}'}{{\Delta_g}^{2}g''}}[({\Delta_g}-2{g'}^{2} ) \alpha_{{1}}+g^2\alpha_8]
    \nonumber\\
    &&-{\frac {{g}^{2}g'\tilde{g}'}{g''{\Delta_g}}}[1+g^2(\alpha_3+\alpha_9)+( g^2-4{g''}^2)\alpha_2]
    -8\,{\frac {\tilde{g}'g'{g}^{2}g''(\beta_{{1}}-\beta_3)}{{\Delta_g}^{2}}}
    \nonumber\\
    &&+\,{\frac {16\tilde{g}'g'{g}^{2}g''}{{\Delta_g}^{3}}}
        [(2g_Z^2-{\Delta_g})\beta_{{1}}\beta_{{3}}-g_Z^2\beta_1^2-4{g''}^2\beta_3^2]
\nonumber\\
C_{+Z'-} &=&\,{\frac {2g''{g}^{2}[\beta_2-( g'^2-4{g''}^2
)\alpha_{{24}}-g'^2\alpha_{25}]}{{\Delta_g}}}
    +\,{\frac {4{g}^{2}g''[g_Z^2\beta_{{1}}\beta_{{2}}-4{g''}^2\beta_2\beta_3]}{{\Delta_g}^{2}}}
    \nonumber\\
    &&-\frac{2}{3}\,{\frac {{g}^{2}g'\tilde{g}'g'' ( 17g_Z^{4}-112g_Z^2{g''}^{2}+32\,{g''}^{4}) {\beta_{{2}}}^{2}}{{\Delta_g}^{3}g_Z^2}}
    -{\frac {{g}^{2}g'\tilde{g}'}{{\Delta_g} g''}}
    -4\,{\frac {g'\tilde{g}'{g}^{2}g''\alpha_{{3}}}{{\Delta_g}}}
    \nonumber\\
    &&-8\,{\frac {g'\tilde{g}'{g}^{2}g''(\beta_{{1}}-\beta_3)}{{\Delta_g}^{2}}}
    -\,{\frac {{g}^{2}g'\tilde{g}'}{{\Delta_g}^{2}g''}}
        [2({g}^{2}{g'}^{2}-2{g''}^{2}{\Delta_g})\alpha_{{1}}+g^2(g'^2-4{g''}^2)\alpha_8]
    \nonumber\\
    &&-16\,{\frac {g'\tilde{g}'{g}^{2}g''}{{\Delta_g}^{3}}}
        [g_Z^2{\beta_{{1}}}^{2}-(2g_Z^2-{\Delta_g})\beta_1\beta_3+4{g''}^2\beta_3^2]
\nonumber\\
D_{+-Z'Z'} &=&4\,{g}^{2}{g''}^{2}(\alpha_{{5}}+\alpha_{21})
    -\,{\frac {4{g}^{4}{g''}^{2}{\beta_{{2}}}^{2}}{{\Delta_g}^{2}}}
    +\,{\frac {4{g}^{4}g'\tilde{g}'}{{\Delta_g}^{2}}}
        [\beta_{{2}}-( g'^2-4{g''}^2)\alpha_{{24}}-g'^2\alpha_{25}]
    \nonumber\\
    &&-\,{\frac {8{g}^{2}g'\tilde{g}'{g''}^{2}\alpha_{{16}}}{{\Delta_g}}}
    +\,{\frac {8{g}^{4}g'\tilde{g}' }{{\Delta_g}^{3}}}
        [(2g_Z^2-{\Delta_g}) \beta_{{1}}\beta_{{2}}-8{g''}^2\beta_2\beta_3]
\nonumber\\
D_{+Z'-Z'} &=&4\,{g}^{2}{g''}^{2}(\alpha_{{4}}+\alpha_{{19}})
    +\,{4\frac {{g}^{4}{g''}^{2}{\beta_{{2}}}^{2}}{{\Delta_g}^{2}}}
    -\,{\frac {4{g}^{4}g'\tilde{g}'}{{\Delta_g}^{2}}}
        [\beta_{{2}}-(g'^2-4{g''}^2)\alpha_{24}-g'^2\alpha_{25}]
    \nonumber\\
    &&-\,{\frac {8{g}^{2}g'\tilde{g}'{g''}^{2}\alpha_{{17}}}{{\Delta_g}}}
    -\,{\frac {8{g}^{4}g'\tilde{g}'}{{\Delta_g}^{3}}}[(2g_Z^{2}-{\Delta_g}) \beta_{{1}}\beta_{{2}}-8{g''}^2\beta_2\beta_3]
\nonumber\\
D_{+-ZZ'} &=&\,{\frac
{4{g}^{4}g''[\beta_2-(g'^2-4{g''}^2)\alpha_{24}-g'^2\alpha_{25}]}{g_Z{\Delta_g}}}
    -2\,{g}^{2}g_Zg''\alpha_{{16}}
    +\,{\frac {8{g}^{4}g''}{{\Delta_g}^2}}(g_Z\beta_{{1}}\beta_{{2}}-4\frac{{g''}^2}{g_Z}\beta_2\beta_3)
    \nonumber\\
    &&-\,{\frac {2{g}^{4}g'\tilde{g}'}{g_Zg''{\Delta_g}}}[1+(2g_Z^2-{\Delta_g})\alpha_3]
-{\frac {g'\tilde{g}'{g}^{6}
(4g'^2{\Delta_g}-12\,{g}^{2}{g''}^{2}+g^2{g_Z}^{2} )
\alpha_{{8}}}{g_Z^3{\Delta_g}^{2}g''}}
    \nonumber\\
    &&-\,{\frac {2g'\tilde{g}'{g}^{4}[g_Z^2(2{g}^{2}g'^2-{g'}^{4}+16{g''}^{4})-4g^2{g''}^2(2{g'}^{2}+g'^2)]\alpha_{{1}}}{g_Z^3{\Delta_g}^{2}g''}}
    \nonumber\\
    &&+8\,{\frac {{g}^{2}g_Zg'\tilde{g}'g''}{{\Delta_g}}}(\alpha_{{7}}-\alpha_{21})
    -16\,{\frac {g'\tilde{g}'{g}^{4}g''}{g_Z{\Delta_g}^{2}}}(\beta_{{1}}-\beta_3)
    -128\,{\frac {g'\tilde{g}'{g}^{4}{g''}^{3}{\beta_{{3}}}^{2}}{g_Z{\Delta_g}^{3}}}
    \nonumber\\
    &&+32\,{\frac {{g}^{4}g'\tilde{g}'g'' }{g_Z{\Delta_g}^{3}}}
        [(2g_Z^2-{\Delta_g}) \beta_{{1}}\beta_{{3}}+\frac{14g_Z^2{g''}^{2}-g_Z^4-4\,{g''}^{4}}{3}\beta_2^2]
    -32\,{\frac {{g}^{4}g'\tilde{g}'g''g_Z{\beta_{{1}}}^{2}}{{\Delta_g}^{3}}}
\nonumber\\
D_{+Z-Z'} &=&D_{+Z'-Z} =\,{\frac
{2{g}^{4}g''[-\beta_2+(g'^2-4{g''}^2)\alpha_{{24}}+g'^2\alpha_{25}]}{g_Z{\Delta_g}}}
    -{g}^{2}g_Zg''\alpha_{{17}}\nonumber\\
    &&-\,{\frac {4{g}^{4}g''}{{\Delta_g}^2}}(g_Z\beta_{{1}}\beta_{{2}}-4\frac{{g''}^2}{g_Z}\beta_2\beta_3)
    +4\,{\frac
{{g}^{2}g_Zg'\tilde{g}'g''}{{\Delta_g}}}(\alpha_{{6}}-\alpha_{19})
    +8\,{\frac {{g}^{4}g'\tilde{g}'g''}{g_Z {\Delta_g}^2}}(\beta_{{1}}-\beta_3)
     \nonumber\\
    &&+{\frac {{g}^{4}g'\tilde{g}'}{g_Zg''{\Delta_g}}}[1+(2g_Z^2-{\Delta_g})\alpha_3]
   +16\,{\frac {{g}^{4}g'\tilde{g}'g''g_Z{\beta_{{1}}}^{2}}{{\Delta_g}^{3}}}
    \nonumber\\
    &&-16\,{\frac {{g}^{4}g'\tilde{g}'g'' }{g_Z{\Delta_g}^{3}}}
        [( {g'}^{2}+4\,{g''}^{2}+{g}^{2} ) \beta_{{1}}\beta_{{3}}-4{g''}^2\beta_3^2]
    +\,{\frac {g'\tilde{g}'{g}^{6} (4g'^2{\Delta_g}-12\,{g}^{2}{g''}^{2}+g^2{g_Z}^{2} ) \alpha_{{8}}}{2g_Z^3{\Delta_g}^{2}g''}}
    \nonumber\\
    &&+{\frac {g'\tilde{g}'{g}^{4} (g^2g_Z^4-4g_Z^{2}{g''}^{2}{\Delta_g}-{g'}^{4}{\Delta_g}) \alpha_{{1}}}{g_Z^3{\Delta_g}^{2}g''}}
    +\,{\frac {16g'\tilde{g}'{g}^{4}g'' ( -14g_Z^2{g''}^{2}+4\,{g''}^{4}+g_Z^4) {\beta_{{2}}}^{2}}{3{\Delta_g}^{3}g_Z^3}}
\nonumber\\
D_{+-AZ'} &=&\,{\frac
{4{g}^{3}g'{g''}[\beta_{{2}}-(g'^2-4{g''}^2)\alpha_{24}-g'^2\alpha_{25}]
}{g_Z{\Delta_g}}}
    -\,{\frac {32g'{g''}{g}^{3}{g''}^{2}\beta_{{2}}\beta_{{3}}}{g_Z{\Delta_g}^2 }}
    +\,{\frac {8{g}^{3}g'{g''}g_Z\beta_{{1}}\beta_{{2}}}{{\Delta_g}^2}}
    \nonumber\\
    &&-\,{\frac {2\tilde{g}'{g}^{3}{g'}^{2}}{{g''}g_Z{\Delta_g}}}
    -\,{\frac {128{g''}^{3}\tilde{g}'{g}^{3}{g'}^{2}{\beta_{{3}}}^{2}}{{\Delta_g}^{3}g_Z}}
    -\,{\frac {4{g}^{3}{g'}^{2}\tilde{g}'{g''}( 17g_Z^{4}-112g_Z^2{g''}^{2}+32\,{g''}^{4}) {\beta_{{2}}}^{2}}{3g_Z^3{\Delta_g}^{3}}}
    \nonumber\\
    &&+\,{\frac {{g'}^{2}\tilde{g}'{g}^{3}}{{\Delta_g}^{2}g_Z^3{g''}}}
        \Big\{[8g_Z^2{g''}^{2}{\Delta_g}-\,{g}^{2}{g'}^{2}(4g_Z^2+2{\Delta_g})]\alpha_1-g^2(3{g'}^{2}{\Delta_g}-8\,{g}^{2}{g''}^{2})\alpha_8\Big\}
    \nonumber\\
    &&-8\,{\frac {{g'}^{2}\tilde{g}'{g''}{g}^{3}\alpha_{{3}}}{g_Z{\Delta_g}}}
    -32\,{\frac {\tilde{g}'{g}^{3}{g''}{g'}^{2}g_Z{\beta_{{1}}}^{2}}{{\Delta_g}^{3}}}
    +32\,{\frac {(2g_Z^2-{\Delta_g})\tilde{g}'{g}^{3}{g''}{g'}^{2}\beta_{{1}}\beta_{{3}}}{{\Delta_g}^{3}g_Z}}
    \nonumber\\
    &&-16\,{\frac {{g'}^{2}\tilde{g}'{g''}{g}^{3}}{{\Delta_g}^{2}g_Z}}(\beta_{{1}}-\beta_3+2\beta_2\beta_3)
\nonumber\\
D_{+A-Z'} &=&D_{+Z'-A}\nonumber\\
 &=&-\,{\frac
{2{g}^{3}g'{g''}[\beta_{{2}}-(g'^2-4{g''}^2)\alpha_{24}-g'^2\alpha_{25}]
}{g_Z{\Delta_g}}}
    +\,{\frac {16g'{g''}{g}^{3}{g''}^{2}\beta_{{2}}\beta_{{3}}}{g_Z{\Delta_g}^2 }}
    -\,{\frac {4{g}^{3}g'{g''}g_Z\beta_{{1}}\beta_{{2}}}{{\Delta_g}^2}}
    \nonumber\\
    &&+\,{\frac {8{g'}^{2}\tilde{g}'{g''}{g}^{3}}{g_Z{\Delta_g}^{2}}}(\beta_{{1}}-\beta_3)
    -\,{\frac {16{g}^{3}{g''}  \tilde{g}'{g'}^{2}}{g_Z{\Delta_g}^{3}}}
        [(2g_Z^2-{\Delta_g})\beta_{{1}}\beta_{{3}}-4{g''}^2\beta_2\beta_3]
    \nonumber\\
    &&+\frac{2}{3}\,{\frac {{g'}^{2}\tilde{g}' ( 17g_Z^{4}-112g_Z^2{g''}^{2}+32\,{g''}^{4}) {g''}{g}^{3}{\beta_{{2}}}^{2}}{g_Z^3{\Delta_g}^{3}}}
    +\,{\frac {16{g}^{3}{g''}g_Z\tilde{g}'{g'}^{2}{\beta_{{1}}}^{2}}{{\Delta_g}^{3}}}
    \nonumber\\
    &&+\,{\frac {4{g'}^{2}\tilde{g}'{g}^{3}{g''}\alpha_{{3}}}{g_Z{\Delta_g}}}
    +{\frac {{g'}^{2}\tilde{g}'{g}^{3} (2g^2g'^2g_Z^2-4g_Z^2{g''}^2{\Delta_g}+\,{g'}^{2}{g}^{2}{\Delta_g}) \alpha_{{1}}}{{g''}g_Z^3{\Delta_g}^{2}}}
    \nonumber\\
    &&-\frac{1}{2}\,{\frac {{g}^{5}{g'}^{2}\tilde{g}' (8\,{g}^{2}{{g''}}^{2}-3g'^2{\Delta_g}) \alpha_{{8}}}{{g''} g_Z^3{\Delta_g}^{2}}}
    +{\frac {{g'}^{2}\tilde{g}'{g}^{3}}{{g''}g_Z{\Delta_g}}}
\nonumber\\
D_{Z'ZZZ} &=&-g_Z^3g''(2\alpha_{15}+\alpha_{{16}}+\alpha_{17})
    \nonumber\\
    &&+4\,{\frac {g'\tilde{g}'g_Z^3g''}{{\Delta_g}}}
        (\alpha_{{6}}+\alpha_7+2\alpha_{10}-2\alpha_{18}-\alpha_{19}-2\alpha_{20}-\alpha_{21})
\nonumber\\
D_{Z'Z'ZZ}
&=&{g''}^{2}g_Z^2(\alpha_{{5}}+2\alpha_7+4\alpha_{20}+2\alpha_{21})
    -4\,{\frac {{g''}^{2}g'\tilde{g}'g_Z^2}{{\Delta_g}}}(2\alpha_{{15}}+\alpha_{16}-2\alpha_{22})
\nonumber\\
D_{Z'ZZ'Z}
&=&4\,{g''}^{2}g_Z^2(\alpha_{{4}}+\alpha_6+2\alpha_{18}+\alpha_{19})
    +\,{32\frac {{g''}^{2}g'\tilde{g}'g_Z^2}{{\Delta_g}}}(\alpha_{{22}}-\alpha_{15})
\nonumber\\
D_{Z'Z'Z'Z}
&=&-4\,{g''}^{3}g_Z(\alpha_{{16}}+\alpha_{17}+2\alpha_{22})\nonumber\\
    &&+16\,{\frac {g'\tilde{g}'{g''}^{3}g_Z}{{\Delta_g}}}
        (\alpha_{{6}}+\alpha_7+2\alpha_{18}-\alpha_{19}+2\alpha_{20}-\alpha_{21}-4\alpha_{23})
\nonumber\\
D_{Z'Z'Z'Z'}
&=&4\,{g''}^{4}(\alpha_{{4}}+\alpha_5+2\alpha_{19}+2\alpha_{21}+4\alpha_{23})
    -\,{\frac
    {16{g''}^{4}g'\tilde{g}'}{{\Delta_g}}}(\alpha_{{16}}+\alpha_{17}+2\alpha_{22})\;.
\label{Z'couplings}
\end{eqnarray}
%%%%%%%%%%%%%%%%%%%%%%%%%%%%%%%%%%%%%%%%%%%%%%%

\end{document}